\documentclass[preprint,showpacs,superscriptaddress,preprintnumbers]{revtex4}
\usepackage{graphicx}

\def\Re{{\cal R}e}
\def\ubar{\bar{u}}
\def\slash#1{{\rlap{\hspace{.08em}/}#1}}

\begin{document}

\title{Two-photon exchange in elastic electron-nucleon scattering}

\author{P.\ G.\ Blunden}
\affiliation{Department of Physics and Astronomy, University of Manitoba,
Winnipeg, MB, Canada R3T 2N2}

\author{W.\ Melnitchouk}
\affiliation{Jefferson Lab, 12000 Jefferson Ave., Newport News, VA 23606}

\author{J.\ A.\ Tjon}
\affiliation{Jefferson Lab, 12000 Jefferson Ave., Newport News, VA 23606}
\affiliation{Department of Physics, University of Maryland, College Park,
MD 20742-4111}

\begin{abstract}
A detailed study of two-photon exchange in unpolarized and polarized
elastic electron--nucleon scattering is presented, taking particular
account of nucleon finite size effects. Contributions from nucleon
elastic intermediate states are found to have a strong angular
dependence, which leads to a partial resolution of the discrepancy
between the Rosenbluth and polarization transfer measurements of the
proton electric to magnetic form factor ratio, $G_E/G_M$.
The two-photon exchange contribution to the longitudinal polarization 
transfer $P_L$ is small, whereas the contribution to the transverse 
polarization transfer $P_T$ is enhanced at backward angles by 
several percent, increasing with $Q^2$. This gives rise to a small,
$\alt 3\%$ suppression of $G_E/G_M$ obtained from the polarization 
transfer ratio $P_T/P_L$ at large $Q^2$.
We also compare the two-photon exchange effects with data on the ratio
of $e^+ p$ to $e^- p$ cross sections, which is predicted to be enhanced
at backward angles.  Finally, we evaluate the corrections to the form 
factors of the neutron, and estimate the elastic intermediate state 
contribution to the $^3$He form factors.
\end{abstract}


\maketitle

\section{Introduction}

Electromagnetic form factors are fundamental observables which
characterize the composite nature of the nucleon. Several decades of
elastic form factor experiments with electron beams, including recent
high-precision measurements at Jefferson Lab and elsewhere, have
provided considerable insight into the detailed structure of the
nucleon.

In the standard one-photon exchange (Born) approximation, the
electromagnetic current operator is parameterized in terms of two form
factors, usually taken to be the Dirac ($F_1$) and Pauli ($F_2$) form
factors,
\begin{eqnarray}
\Gamma^\mu
= F_1(q^2)\ \gamma^\mu
+ {i \sigma^{\mu\nu} q_\nu \over 2 M} F_2(q^2)\ ,
\label{eq:current}
\end{eqnarray}
where $q$ is the momentum transfer to the nucleon, and $M$ is the
nucleon mass.
The resulting cross section depends on two kinematic variables,
conventionally taken to be $Q^2 \equiv -q^2$ (or $\tau\equiv Q^2/4M^2$)
and either the scattering angle $\theta$, or the virtual photon
polarization $\varepsilon = \left(1 + 2 (1+\tau)
\tan^2{(\theta/2)}\right)^{-1}$. In terms of the Sachs electric and
magnetic form factors, defined as
\begin{eqnarray}
G_E(Q^2) &=& F_1(Q^2) - \tau F_2(Q^2)\ ,	\\
G_M(Q^2) &=& F_1(Q^2) + F_2(Q^2)\ ,
\end{eqnarray}
the reduced Born cross section can be written
\begin{eqnarray}
\sigma_R = G_M^2(Q^2) + { \varepsilon \over \tau} G_E^2(Q^2)\ .
\label{eq:sigmaR}
\end{eqnarray}

The standard method which has been used to determine the electric and
magnetic form factors, particularly those of the proton, has been the
Rosenbluth, or longitudinal-transverse (LT), separation method. Since
the form factors in Eq.~(\ref{eq:sigmaR}) are functions of $Q^2$ only,
studying the cross section as a function of the polarization
$\varepsilon$ at fixed $Q^2$ allows one to extract $G_M^2$ from the
$\varepsilon$-intercept, and the ratio $R \equiv \mu G_E/G_M$ from the
slope in $\varepsilon$, where $\mu$ is the nucleon magnetic moment.
The results of the Rosenbluth measurements for the proton have generally
been consistent with $R \approx 1$ for $Q^2 \alt 6$~GeV$^2$
\cite{Wal94,Arr03,Chr04}. The ``Super-Rosenbluth'' experiment at
Jefferson Lab \cite{Qat04}, in which smaller systematic errors were
achieved by detecting the recoiling proton rather than the electron, as
in previous measurements, is also consistent with the earlier LT
results.

An alternative method of extracting the ratio $R$ has been developed
recently at Jefferson Lab~\cite{Jon00}, in which a polarized electron
beam scatters from an unpolarized target, with measurement of the
polarization of the recoiling proton. From the ratio of the transverse
to longitudinal recoil polarizations one finds
\begin{eqnarray}
R &=& -\mu { E_1+E_3\over 2 M }
     \tan{\theta\over 2}\, { P_T \over P_L } 
= -\mu \sqrt{\tau (1+\varepsilon)\over 2 \varepsilon}\, {P_T \over P_L }, 
\,
\label{eq:poltrans}
\end{eqnarray}
where $E_1$ and $E_3$ are the initial and final electron energies, and
$P_T$ ($P_L$) is the polarization of the recoil proton transverse
(longitudinal) to the proton momentum in the scattering plane.
The polarization transfer experiments yielded strikingly different
results compared with the LT separation, with $R \approx 1-0.135
(Q^2/{\rm GeV}^2 - 0.24)$ over the same range in $Q^2$ \cite{Arr03}.
Recall that in perturbative QCD one expects $F_1 \sim Q^2 F_2$ at large
$Q^2$ (or equivalently $G_E \sim G_M$) \cite{pQCD}, so that these
results imply a strong violation of scaling behavior (see also
Refs.~\cite{Ral03,Bel03}).

The question of which experiments are correct has been debated over the
past several years. Attempts to reconcile the different measurements
have been made by several authors \cite{Gui03,Blu03,Che04,Afa05}, who
considered whether 2$\gamma$ exchange effects, which form part of the
radiative corrections (RCs), and which are treated in an approximate
manner in the standard RC calculations \cite{MT69}, could account for
the observed discrepancy. An explicit calculation \cite{Blu03} of the
two-photon exchange diagram, in which nucleon structure effects were for
the first time fully incorporated, indeed showed that around half of the
discrepancy could be removed just by the nucleon elastic intermediate
states. A partonic level calculation \cite{Che04,Afa05} subsequently
showed that the deep inelastic region can also contribute significantly 
to the box diagram.

In this paper we further develop the methodology introduced in
Ref.~\cite{Blu03}, and apply it to systematically calculate the
2$\gamma$ exchange effects in a number of electron--nucleon scattering
observables. We focus on the nucleon elastic intermediate states;
inelastic contributions are discussed elsewhere \cite{Kon05}. In Sec.~II we 
examine the effects of 2$\gamma$ exchange on the ratio of electric to 
magnetic form factors in unpolarized scattering. In contrast to the 
earlier analysis \cite{Blu03}, in which simple monopole form factors were 
utilized at the internal $\gamma NN$ vertices, here we parameterize the 
vertices by realistic form factors, and study the model-dependence of 
effects on the ratio $R$ due to the choice of form factors. We also 
compare the results with data on the ratio of $e^+ p$ to $e^- p$ 
scattering cross sections, which is directly sensitive to 2$\gamma$ 
exchange effects.

In Sec.~III we examine the effects of 2$\gamma$ exchange on the
polarization transfer reaction, $\vec e p \to e \vec p$, for both
longitudinally and transversely polarized recoil protons. We also
consider the case of proton polarization normal to the reaction plane,
which depends on the imaginary part of the box diagram. Since this is
absent in the Born approximation, the normal polarization provides a
clean signature of 2$\gamma$ exchange effects, even though it does not
directly address the $G_E^p/G_M^p$ discrepancy.
Following the discussion of the proton, in Sec.~IV we consider 2$\gamma$
exchange corrections to the form factors of the neutron, both for the LT
separation and polarization transfer techniques. Applying the same
formalism to the case of the $^3$He nucleus, in Sec.~V we compute the
elastic contribution from the box diagram to the ratio of charge to
magnetic form factors of $^3$He. In Sec.~VI we summarize our findings,
and discuss future work.

\section{Two-photon exchange in unpolarized scattering}

In this section we outline the formalism used to calculate the 2$\gamma$
exchange contribution to the unpolarized electron--nucleon cross
section, and examine the effect on the $G_E^p/G_M^p$ ratio extracted
using LT separation. Since there are in general three form factors that
are needed to describe elastic $e N$ scattering beyond 1$\gamma$
exchange, we also evaluate the 2$\gamma$ contributions to each of the
form factors separately. In the final part of this section, we examine
the effect of the 2$\gamma$ correction on the ratio of $e^+ p$ to $e^-
p$ elastic cross sections, which is directly sensitive to 2$\gamma$
exchange effects.

\subsection{Formalism}

For the elastic scattering process we define the momenta of the initial
electron and nucleon as $p_1$ and $p_2$, and of the final electron and
nucleon as $p_3$ and $p_4$, respectively, $e(p_1) + p(p_2) \to e(p_3) +
p(p_4)$. The four-momentum transferred from the electron to the nucleon
is given by $q = p_4-p_2 = p_1-p_3$ (with $Q^2 \equiv -q^2 > 0$), and
the total electron and proton invariant mass squared is given by $s =
(p_1+p_2)^2 = (p_3+p_4)^2$.
In the Born approximation, the amplitude can be written
\begin{equation}
{\cal M}_0 = -i {e^2\over q^2} \ubar(p_3) \gamma_\mu u(p_1)\ 
\ubar(p_4) \Gamma^\mu (q) u(p_2)\ ,
\label{eq:m0}
\end{equation}
where $e$ is the electron charge, and $\Gamma^\mu$ is given by
Eq.~(\ref{eq:current}).
In terms of the amplitude ${\cal M}_0$, the corresponding differential 
Born cross section is given by
\begin{eqnarray}
{ d\sigma_0 \over d\Omega }
&=& \left( { \alpha \over 4 M q^2 } {E_3 \over E_1} \right)^2
    \left| {\cal M}_0 \right|^2\ 
= \sigma_{\rm Mott}
    { \tau \over \varepsilon\ (1+\tau) }\
    \sigma_R\ ,
\label{eq:sigma0}
\end{eqnarray}
where $\sigma_R$ is the reduced cross section given in
Eq.~(\ref{eq:sigmaR}), and the Mott cross section for the scattering
from a point particle is
\begin{eqnarray}
\sigma_{\rm Mott}
&=& { \alpha^2 E_3 \cos^2{\theta \over 2} \over
      4 E_1^3 \sin^4{\theta \over 2} }\ ,
\end{eqnarray}
with $E_1$ and $E_3$ the initial and final electron energies, and
$\alpha = e^2/4\pi$ the electromagnetic fine structure constant.
Including radiative corrections to order $\alpha$, the elastic
scattering cross section is modified as
\begin{eqnarray}
{d\sigma_0 \over d\Omega}
&\to& {d\sigma \over d\Omega} (1 + \delta)\ , 
\end{eqnarray}
where $\delta$ includes one-loop virtual corrections
(vacuum polarization, electron and proton vertex, and two photon
exchange corrections), as well as inelastic bremsstrahlung for real
photon emission \cite{MT69}.

According to the LT separation technique, one extracts the ratio $R^2$
from the $\varepsilon$ dependence of the cross section at fixed $Q^2$.
Because of the factor $\varepsilon/\tau$ multiplying $G_E^2$ in
Eq.~(\ref{eq:sigmaR}), the cross section becomes dominated by $G_M^2$ with
increasing $Q^2$, while the relative contribution of the $G_E^2$ term is
suppressed. Hence understanding the $\varepsilon$ dependence of the 
radiative correction $\delta$ becomes increasingly important at high $Q^2$.
As pointed out in Ref.~\cite{Arr03}, for example, a few percent change
in the $\varepsilon$ slope in $d\sigma$ can lead to a sizable effect on
$R$. In contrast, as we discuss in Sec.~III below, the polarization
transfer technique does not show the same sensitivity to the
$\varepsilon$ dependence of $\delta$.

If we denote the amplitude for the one-loop virtual corrections by
${\cal M}_1$, then ${\cal M}_1$ can be written as the sum of a
``factorizable'' term, proportional to the Born amplitude ${\cal M}_0$,
and a non-factorizable part $\overline{\cal M}_1$,
\begin{equation}
{\cal M}_1 = f(Q^2,\varepsilon) {\cal M}_0 + \overline{\cal M}_1\ .
\label{eq:m1}
\end{equation}
The ratio of the full cross section (to order $\alpha$) to the Born
can therefore be written as
\begin{eqnarray}
1 + \delta
&=& { \left| {\cal M}_0 + {\cal M}_1 \right|^2 \over
      \left| {\cal M}_0 \right|^2 }\ ,
\end{eqnarray}
with $\delta$ given by
\begin{equation}
\delta = 2 f(Q^2,\varepsilon)
       + { 2 \Re\{{\cal M}_0^\dagger \overline{\cal M}_1\}
	   \over |{\cal M}_0|^2}\ .
\label{eq:delta}
\end{equation}
In practice the factorizable terms parameterized by
$f(Q^2,\varepsilon)$, which includes the electron vertex correction,
vacuum polarization, and the infrared (IR) divergent parts of the
nucleon vertex and two-photon exchange corrections, are found to be
dominant. Furthermore, these terms are all essentially independent of
hadronic structure.

However, as explained in Ref.~\cite{Blu03}, the contributions to the
functions $f(Q^2,\varepsilon)$ from the electron vertex, vacuum
polarization, and proton vertex terms depend only on $Q^2$, and
therefore have no relevance for the LT separation aside from an overall
normalization factor. Hence, of the factorizable terms, only the IR
divergent two-photon exchange contributes to the $\varepsilon$
dependence of the virtual photon corrections.

The terms which do depend on hadronic structure are contained in
$\overline{\cal M}_1$, and arise from the finite nucleon vertex and
two-photon exchange corrections. For the case of the proton, the
hadronic vertex correction was analyzed by Maximon and Tjon~\cite{MT00},
and found to be $< 0.5\%$ for $Q^2 < 6$~GeV$^2$. Since the proton vertex
correction does not have a strong $\varepsilon$ dependence, it will not
affect the LT analysis, and can be safely neglected.

\begin{figure}
\includegraphics[width=14cm]{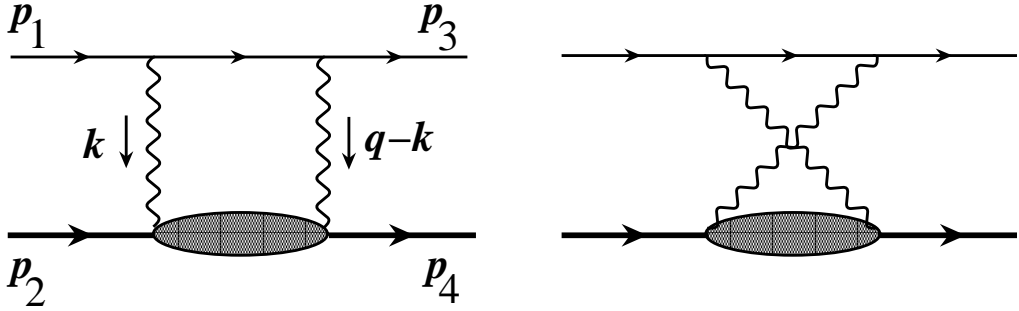}
\vspace*{-1cm}
\caption{Two-photon exchange box and crossed box diagrams
	for elastic electron--proton scattering.
\label{fig:diag}}
\end{figure}

For the inelastic bremsstrahlung cross section, the amplitude for real
photon emission can also be written in the form of Eq.~(\ref{eq:m1}). In
the soft photon approximation the amplitude is completely factorizable.
A significant $\varepsilon$ dependence arises due to the
frame dependence of the angular distribution of the emitted photon.
These corrections, together with external bremsstrahlung, contain the
main $\varepsilon$ dependence of the radiative corrections, and are
usually accounted for in the experimental analyses. They are generally
well understood, and in fact enter differently depending on whether
the electron or proton are detected in the final state. Hence
corrections beyond the standard ${\cal O}(\alpha)$ radiative corrections
which can lead to non-negligible $\varepsilon$ dependence are confined
to the 2$\gamma$ exchange diagrams, illustrated in Fig.~\ref{fig:diag},
and are denoted by ${\cal M}^{2\gamma}$, which we will focus on in the
following. The 2$\gamma$ exchange correction $\delta^{2\gamma}$ which we
calculate is then essentially
\begin{eqnarray}
\delta^{2\gamma} &\to&
{ 2 \Re \left\{ {\cal M}_0^\dagger {\cal M}^{2\gamma} \right\}
  \over \left| {\cal M}_0 \right|^2}\ .
\label{eq:delta_eff}
\end{eqnarray}

In principle the two-photon exchange amplitude ${\cal M}^{2\gamma}$
includes all possible hadronic intermediate states in
Fig.~\ref{fig:diag}. Here we consider only the elastic contribution to
the full response function, and assume that the proton propagates as a
Dirac particle (excited state contributions are considered in
Ref.~\cite{Kon05}). We also assume that the structure of the off-shell
current operator is similar to that in Eq.~(\ref{eq:current}), and use
phenomenological form factors at the $\gamma NN$ vertices. This is of
course the source of the model dependence in the problem. Clearly this
also creates a tautology, as the radiative corrections are also used to
determine the experimental form factors. However, because $\delta$ is a
ratio, the model dependence cancels somewhat, provided the same
phenomenological form factors are used for both ${\cal M}_0$ and ${\cal
M}^{2\gamma}$ in Eq.~(\ref{eq:delta_eff}).

The total 2$\gamma$ exchange amplitude, including the box and crossed
box diagrams in Fig.~\ref{fig:diag}, has the form
\begin{eqnarray}
{\cal M}^{2\gamma}
&=& e^4 \int {d^4 k\over (2\pi)^4}
    {N_{\rm box}(k) \over D_{\rm box}(k)}
	+ e^4 \int {d^4 k\over (2\pi)^4}
    {N_{\rm x-box}(k) \over D_{\rm x-box}(k)}\ ,
\label{eq:mbox}
\end{eqnarray}
where the numerators are the matrix elements
\begin{eqnarray}
N_{\rm box}(k)
&=& \ubar(p_3) \gamma_\mu (\slash{p}_1 - \slash{k} + m)
    \gamma_\nu u(p_1)				\nonumber\\
&\times& \ubar(p_4) \Gamma^\mu(q-k) (\slash{p}_2 + \slash{k} + M)
	 \Gamma^\nu(k) u(p_2)\ ,
\label{eq:nbox}					\\
N_{\rm x-box}(k)
&=& \ubar(p_3) \gamma_\nu (\slash{p}_3 + \slash{k} + m)
    \gamma_\mu u(p_1)				\nonumber\\
&\times& \ubar(p_4) \Gamma^\mu(q-k) (\slash{p}_2 + \slash{k} + M)
	 \Gamma^\nu(k) u(p_2)\ ,
\label{eq:nxbox}
\end{eqnarray}
and the denominators are products of propagators
\begin{eqnarray}
D_{\rm box}(k)
&=& [k^2-\lambda^2] [(k-q)^2-\lambda^2]		\nonumber\\
&&\times [(p_1-k)^2-m^2] [(p_2+k)^2-M^2]\ ,
\label{eq:dbox}					\\
D_{\rm x-box}(k)
&=& \left. D_{\rm box}(k)\right|_{p_1-k\rightarrow p_3+k}\ .
\label{eq:dxbox}
\end{eqnarray}
An infinitesimal photon mass $\lambda$ has been introduced in the photon
propagator to regulate the IR divergences. The IR divergent part is of
interest since it is the one usually included in the standard RC
analyses. The finite part, which is typically neglected, has been
included in Ref.~\cite{Blu03} and found to have significant $\varepsilon$
dependence.

The IR divergent part of the amplitude ${\cal M}^{2\gamma}$ can be
separated from the IR finite part by analyzing the structure of the
photon propagators in the integrand of Eq.~(\ref{eq:mbox}). The two
poles, where the photons are soft, occur at $k=0$ and at $k=q$. The
dominant (IR divergent) contribution to the integral (\ref{eq:mbox})
comes from the poles, and one therefore typically makes the
approximation
\begin{equation}
{\cal M}_{\rm IR}^{2\gamma} \approx
e^4 N_{\rm box}(0) \int {d^4 k\over (2\pi)^4}
{1\over D_{\rm box}(k)} + 
e^4 N_{\rm x-box}(0) \int {d^4 k\over (2\pi)^4}
{1\over D_{\rm x-box}(k)}\ ,\label{eq:mboxapp}
\end{equation}
with
\begin{eqnarray}
N_{\rm box}(q)&=&N_{\rm box}(0)=4 i p_1\cdot p_2 {q^2{\cal M}_0\over e^2}\ ,\\
N_{\rm x-box}(q)&=&N_{\rm x-box}(0)=4 i p_3\cdot p_2 {q^2{\cal M}_0\over
e^2}\ .
\end{eqnarray}
In this case the IR divergent contribution is proportional to the Born
amplitude, and the corresponding correction to the Born cross section is
independent of hadronic structure.

The remaining integrals over propagators can be done analytically. In
the target rest frame the total IR divergent two-photon exchange
contribution to the cross section is found to be
\begin{equation}
\delta_{\rm IR}
= - {2 \alpha\over \pi} \ln{\left(E_1\over E_3\right)}
  \ln{\left(Q^2\over\lambda^2\right)}\ ,
\label{eq:deltaIR}
\end{equation}
a result given by Maximon and Tjon~\cite{MT00}.
The logarithmic IR singularity in $\lambda$ is exactly cancelled by a corresponding
term in the bremsstrahlung cross section involving the interference
between real photon emission from the electron and from the nucleon.

By contrast, in the standard treatment of Mo and Tsai (MT)~\cite{MT69}
a different approximation for the integrals over propagators is introduced.
Here, the IR divergent contribution to the cross section is
\begin{equation}
\delta_{\rm IR}({\rm MT})
= -2{\alpha\over \pi} \left( K(p_1,p_2) - K(p_3,p_2)\right)\ ,
\label{eq:deltaIRMT}
\end{equation}
where $K(p_i,p_j) = p_i\cdot p_j\,\int_0^1 dy\,\ln{(p_y^2/
\lambda^2)}/p_y^2$ and $p_y=p_i y + p_j (1-y)$. The logarithmic
dependence on $\lambda$ is the same as Eq.\ (\ref{eq:deltaIR}), however.

As mentioned above, the full expression in Eq.~(\ref{eq:mbox}) includes
both finite and IR divergent terms, and form factors at the $\gamma NN$
vertices. In Ref.~\cite{Blu03} the proton form factors $F_1$ and $F_2$
were expressed in terms of the Sachs electric and magnetic form factors,
\begin{eqnarray}
F_1(Q^2) &=& {G_E(Q^2) + \tau G_M(Q^2) \over 1 + \tau}\ , \\
F_2(Q^2) &=& {G_M(Q^2) - G_E(Q^2) \over 1 + \tau}\ ,
\end{eqnarray}
with $G_E$ and $G_M$ both parameterized by a simple monopole form,
$G_{E,M}(Q^2) \sim \Lambda^2/(\Lambda^2+Q^2)$, with the mass parameter
$\Lambda$ related to the size of the proton.
In the present analysis we generalize this approach by using more
realistic form factors in the loop integration, consistent with the
actual $G_{E,M}$ data. The functions $F_1$ and $F_2$ are parameterized
directly in terms of sums of monopoles, of the form
\begin{eqnarray}
F_{1,2}(Q^2)
&=& \sum_{i=1}^N { n_i \over d_i + Q^2 }\ ,
\label{eq:3pole}
\end{eqnarray}
where $n_i$ and $d_i$ are free parameters, and $n_N$ is determined
from the normalization condition,
$n_N = d_N (F_{1,2}(0) - \sum_{i=1}^{N-1} n_i/d_i)$.
The parameters $n_i$ and $d_i$ for the $F_1$ and $F_2$
form factors of the proton and neutron are given in Table~I.
The normalization conditions are $F_1^p(0) = 1$ and
$F_2^p(0) = \kappa_p$ for the proton, and
$F_1^n(0) = 0$ and $F_2^n(0) = \kappa_n$ for the neutron,
where $\kappa_p = 1.793$ and $\kappa_n = -1.913$ are the
proton and neutron anomalous magnetic moments, respectively.

In practice we use the parameterization from Ref.~\cite{Mer96},
and fit the parameterized form factors a sum of three monopoles,
except for $F_2^n$, which is fitted with $N=2$.
As discussed in the next section, the sensitivity of the results
to the choice of form factor is relatively mild.
Of course, one should note that the data to which the form factors
are fitted were extracted under the assumption of 1$\gamma$ exchange,
so that in principle one should iterate the data extraction and
fitting procedure for self-consistency.
However, within the accuracy of the data and of the 2$\gamma$
calculation the effect of this will be small.

To obtain the radiatively corrected cross section for unpolarized
electron scattering the polarizations of the incoming and outgoing
electrons and nucleons in Eqs.~(\ref{eq:nbox}) and (\ref{eq:nxbox}) need
to be averaged and summed, respectively. The resulting expression
involves a product of traces in the Dirac spaces of the electron and
nucleon. The trace algebra is tedious but straightforward. It was
carried out using the algebraic program FORM \cite{Vermaseren} and
verified independently using the program Tracer \cite{Tracer}. We also
used two independent Mathematica packages (FeynCalc \cite{feyncalc} and
FormCalc \cite{formcalc}) to carry out the loop integrals. The packages
gave distinct but equivalent analytic expressions, which gave identical
numerical results. The loop integrals in Eq.~(\ref{eq:mbox}) can be
expressed in terms of four-point Passarino-Veltman functions
\cite{PV79}, which have been calculated using Spence function
\cite{HV79} as implemented by Veltman \cite{Veltman}. In the actual
calculations we have used the FF program \cite{ff}. The results of the
proton calculation are presented in the following section.

\subsection{2$\gamma$ Corrections to Proton Form Factors}

In typical experimental analyses of electromagnetic form factor data
\cite{Wal94} radiative corrections are implemented using the
prescription of Ref.~\cite{MT69}, including using
Eq.~(\ref{eq:deltaIRMT}) to approximate the 2$\gamma$ contribution. To
investigate the effect of our results on the data analyzed in this
manner, we will therefore compare the $\varepsilon$ dependence of the
full calculation with that of $\delta_{\rm IR}({\rm MT})$. To make the
comparison meaningful, we will consider the difference
\begin{eqnarray}
\Delta &\equiv& \delta_{\rm full} - \delta_{\rm IR}({\rm MT})\ ,
\label{eq:Delta_dif}
\end{eqnarray}
in which the IR divergences cancel, and which is independent of
$\lambda$.

\begin{figure}[h]
\includegraphics[height=12cm,angle=270]{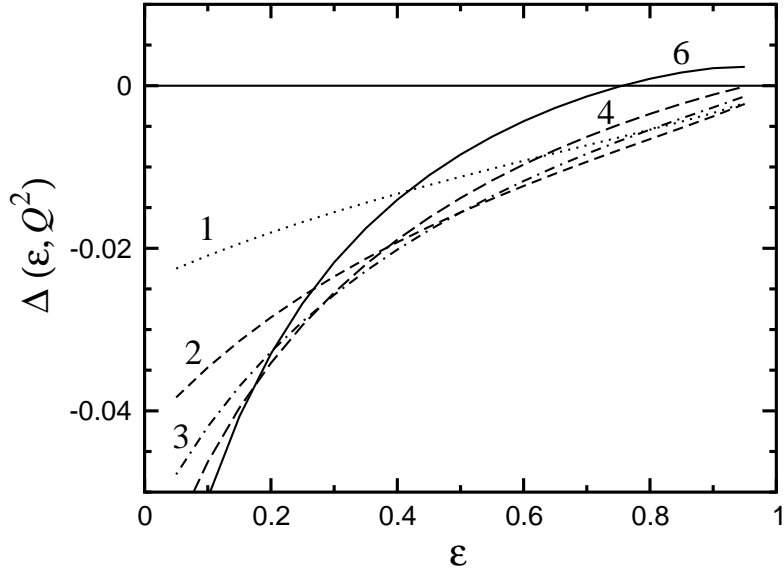}
\caption{Difference between the full two-photon exchange correction
	to the elastic cross section (using the realistic form factors
	in Eq.~(\protect\ref{eq:3pole})) and the commonly used
	expression (\ref{eq:deltaIRMT}) from Mo \& Tsai \cite{MT69}
	for $Q^2 = 1$--6~GeV$^2$.
	The numbers labeling the curves denote the respective $Q^2$
	values in GeV$^2$.
\label{fig:Delta}}
\end{figure}

The results for the difference $\Delta$ between the full calculation and
the MT approximation are shown in Fig.~\ref{fig:Delta} for several
values of $Q^2$ from 1 to 6~GeV$^2$. The additional corrections are most
significant at low $\varepsilon$, and essentially vanish at large
$\varepsilon$. At the lower $Q^2$ values $\Delta$ is approximately
linear in $\varepsilon$, but significant deviations from linearity are
observed with increasing $Q^2$, especially at smaller $\varepsilon$.

\begin{figure}[h]
\includegraphics[height=12cm,angle=270]{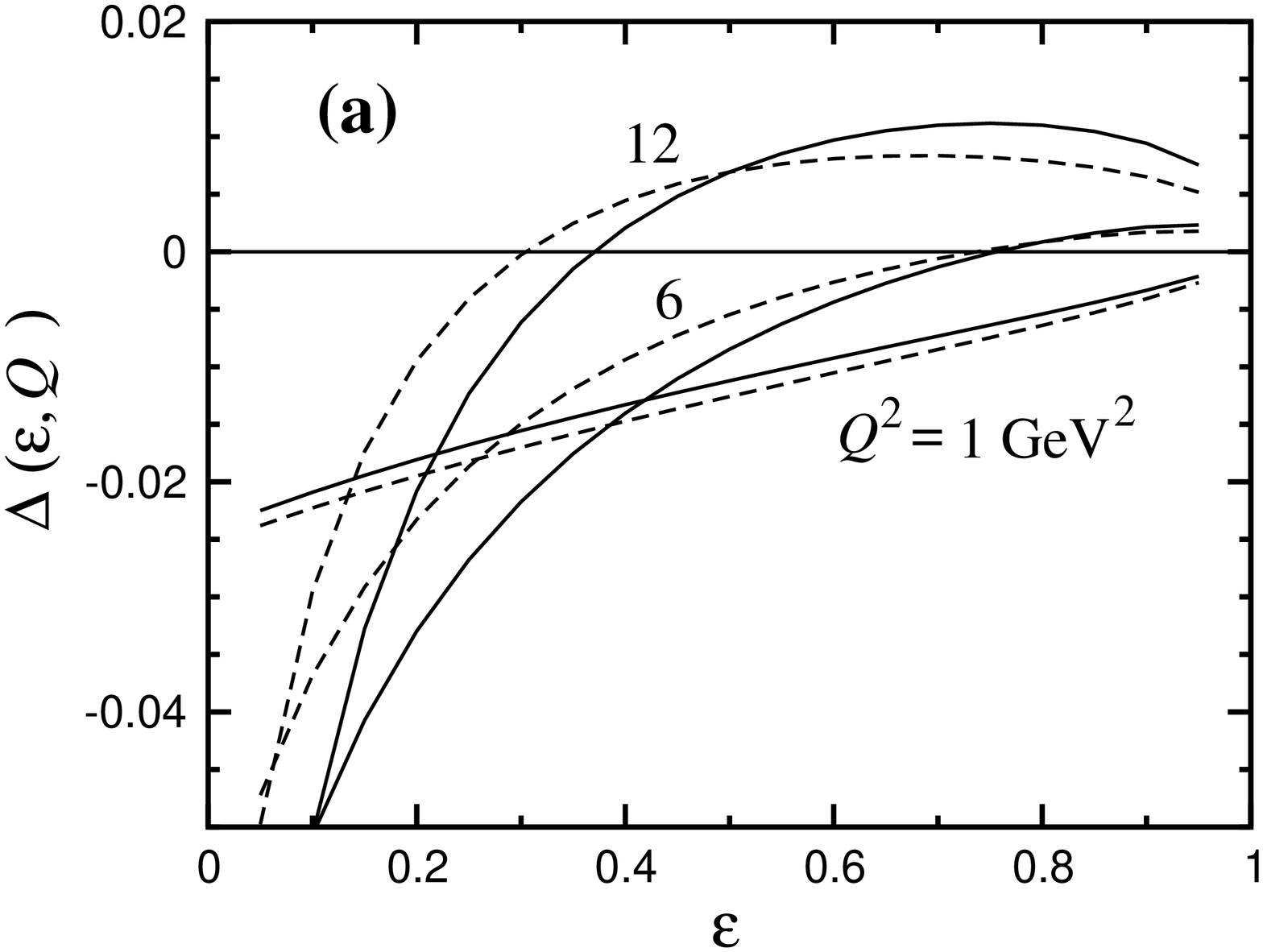}
\includegraphics[height=12cm,angle=270]{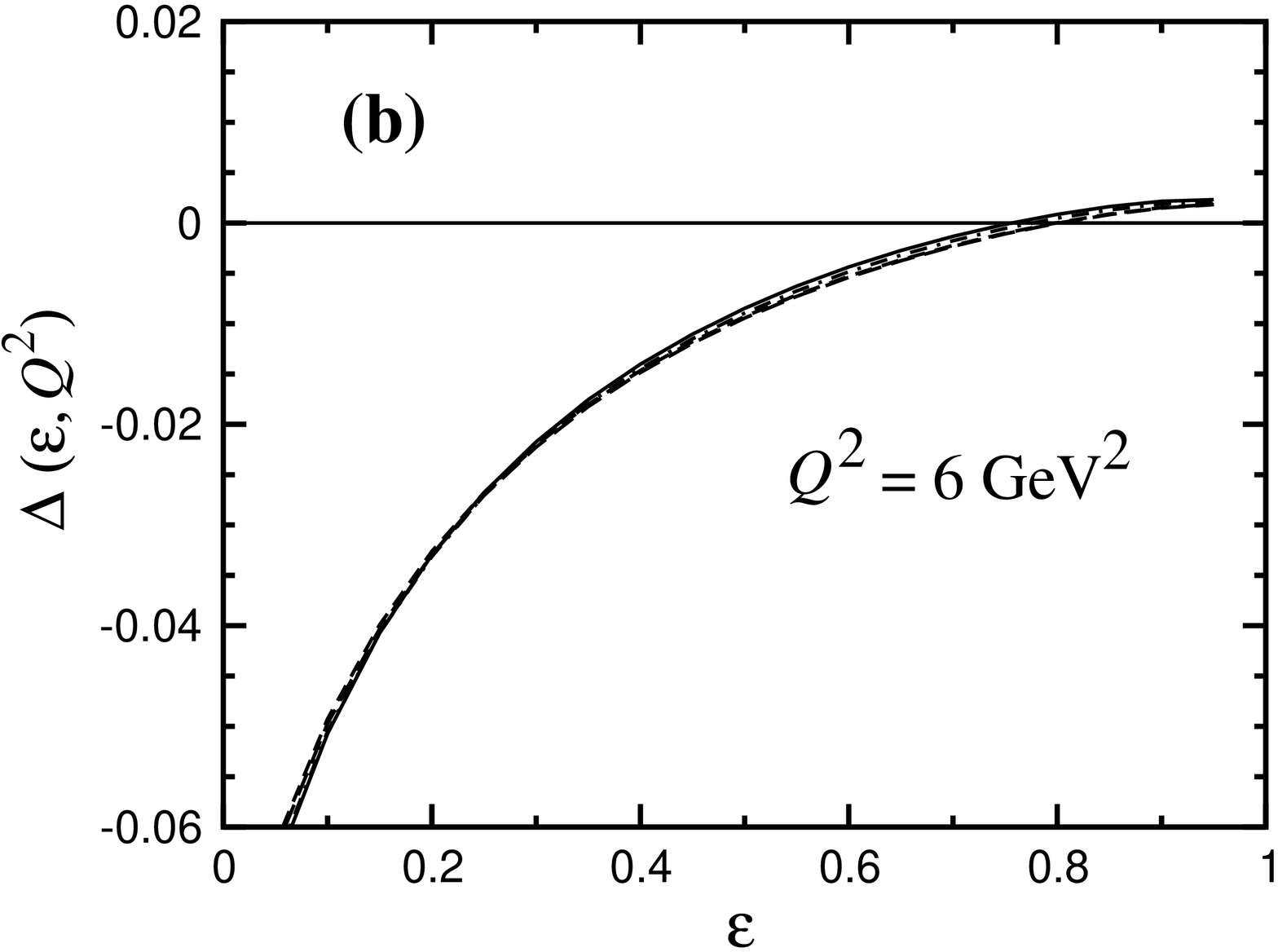}
\caption{Model dependence of the difference between the full
	two-photon exchange correction and the Mo \& Tsai approximation:
    (a) at $Q^2 = 1$, 6 and 12~GeV$^2$, using realistic (solid)
	\cite{Mer96} and dipole (dashed) form factors;
    (b) at $Q^2 = 6$~GeV$^2$ using the form factor parameterizations
	from Refs.~\cite{Mer96} (solid), \cite{Bra02} (dashed), and
	\cite{Arr04} with $G_E^p$ constrained by the LT-separated
	(dot-dashed) and polarization transfer (long-dashed) data.
\label{fig:ffcomp}}
\end{figure}

In Fig.~\ref{fig:ffcomp}(a) we illustrate the model dependence of the
results by comparing the results in Fig.~\ref{fig:Delta} at $Q^2 = 1$
and 6~GeV$^2$ with those obtained using a dipole form for the $F_1^p$
and $F_2^p$ form factors, with mass $\Lambda = 0.84$~GeV. At the lower,
$Q^2=1$~GeV$^2$, value the model dependence is very weak, with
essentially no change at all in the slope. For the larger value
$Q^2=6$~GeV$^2$ the differences are slightly larger, but the general
trend of the correction remains unchanged. We can conclude therefore
that the model dependence of the calculation is quite modest. Also
displayed is the correction at $Q^2=12$~GeV$^2$, which will be accessible
in future experiments, showing significant deviations from linearity over 
the entire $\varepsilon$ range.

The results are also relatively insensitive to the high-$Q^2$ behavior
of the $G_E^p/G_M^p$ ratio, as Fig.~\ref{fig:ffcomp}(b) illustrates.
Here the correction $\Delta$ is shown at $Q^2=6$~GeV$^2$ calculated
using various form factor inputs, from parameterizations obtained by
fitting only the LT-separated data \cite{Mer96,Arr04}, and those in
which $G_E^p$ is constrained by the polarization transfer data
\cite{Arr04,Bra02}. The various curves are almost indistinguishable, and
the dependence on the form factor inputs at lower $Q^2$ is expected to
be even weaker than that in Fig.~\ref{fig:ffcomp}(b).

\begin{figure}[h]
\includegraphics[height=12cm,angle=270]{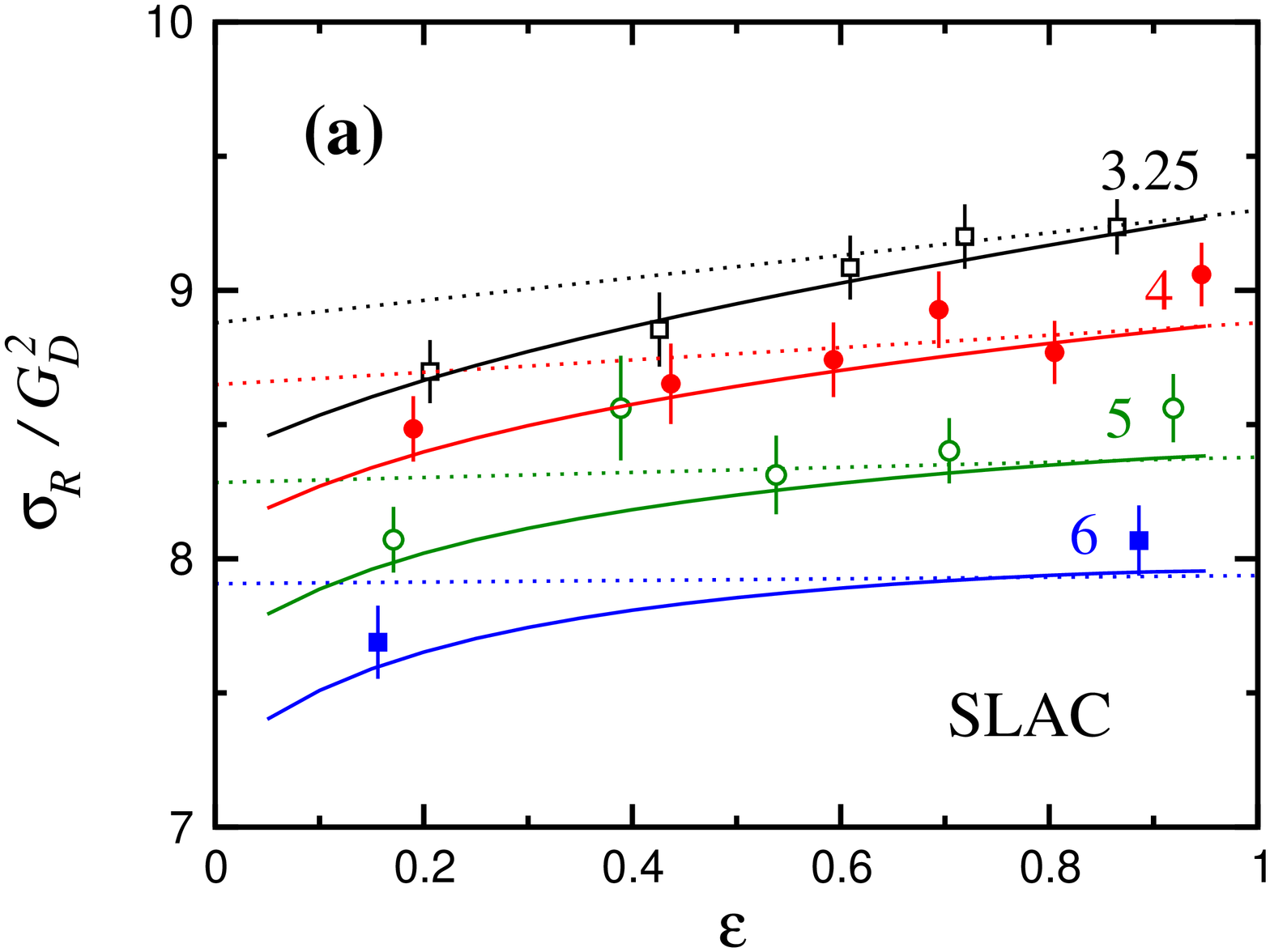}
\includegraphics[height=12cm,angle=270]{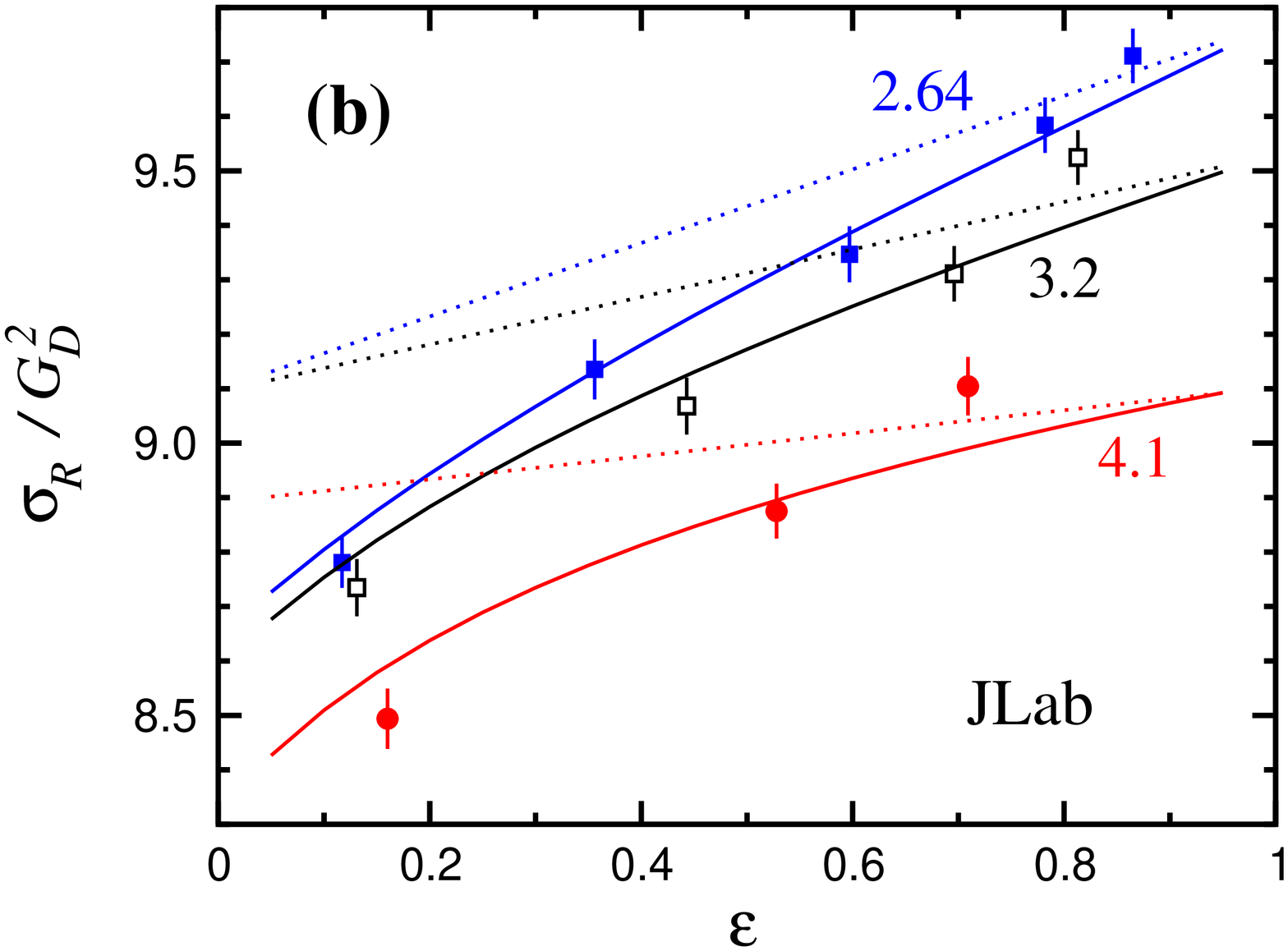}
\caption{Reduced cross section $\sigma_R$ (scaled by the dipole form
	factor $G_D^2$) versus $\varepsilon$ for several values of $Q^2$:
    (a) SLAC data \cite{And94} at $Q^2=3.25$ (open squares),
	4 (filled circles), 5 (open circles) and 6~GeV$^2$
	(filled squares);
    (b) JLab data \cite{Qat04} at $Q^2=2.64$ (filled squares),
	3.2 (open squares) and 4.1~GeV$^2$ (filled circles).
	The dotted curves are Born cross sections evaluated using a
	form factor parameterization \cite{Bra02} with $G_E^p$ fitted
	to the polarization transfer data \cite{Jon00},
	while the solid curves include 2$\gamma$ contributions.
	The curves in the bottom panel have been shifted by
	(+1.0\%, +2.1\%, +3.0\%) for $Q^2=(2.64, 3.2, 4.1)$~GeV$^2$.
\label{fig:sigR}}
\end{figure}

The effect of the 2$\gamma$ corrections on the cross sections can be
seen in Fig.~\ref{fig:sigR}, where the reduced cross section $\sigma_R$,
scaled by the square of the dipole form factor,
\begin{eqnarray}
G_D &=& \left( 1 + {Q^2 \over 0.71\ {\rm GeV}^2} \right)^{-2}\ ,
\label{eq:Gdip}
\end{eqnarray}
is plotted as a function of $\varepsilon$ for several fixed values of
$Q^2$. In Fig.~\ref{fig:sigR}(a) the results are compared with the SLAC
data \cite{And94} at $Q^2=3.25$, 4, 5 and 6~GeV$^2$, and with data from
the ``Super-Rosenbluth'' experiment at JLab \cite{Qat04} in
Fig.~\ref{fig:sigR}~(b). In both cases the Born level results (dotted
curves), which are obtained using the form factor parameterization of
Ref.~\cite{Bra02} in which $G_E^p$ is fitted to the polarization
transfer data \cite{Jon00}, have slopes which are significantly
shallower than the data. With the inclusion of the 2$\gamma$
contribution (solid curves), there is a clear increase of the slope,
with some nonlinearity evident at small $\varepsilon$. The corrected
results are clearly in better agreement with the data, although do not
reproduce the entire correction necessary to reconcile the Rosenbluth
and polarization transfer measurements.

To estimate the influence of these corrections on the electric to
magnetic proton form factor ratio, the simplest approach is to examine
how the $\varepsilon$ slope changes with the inclusion of the 2$\gamma$
exchange. Of course, such a simplified analysis can only be approximate
since the $\varepsilon$ dependence is only linear over limited regions
of $\varepsilon$, with clear deviations from linearity at low
$\varepsilon$ and high $Q^2$. In the actual data analyses one should
apply the correction $\Delta$ directly to the data, as in
Fig.~\ref{fig:sigR}. However, it is still instructive to obtain an
estimate of the effect on $R$ by taking the slope over several ranges of
$\varepsilon$.

Following Ref.~\cite{Blu03}, this can be done by fitting the correction
$(1+\Delta)$ to a linear function of $\varepsilon$, of the form $a +
b\varepsilon$, for each value of $Q^2$ at which the ratio $R$ is
measured. The corrected reduced cross section in Eq.~(\ref{eq:sigmaR})
then becomes
\begin{equation}
\sigma_R
\approx a\ G_M^2(Q^2)
\left[ 1 + {\varepsilon \over \mu^2 \tau}
	   \left( R^2 \left[ 1 + \varepsilon b/a \right]
		+ \mu^2 \tau b/a
	   \right)
\right]\ ,
\label{eq:sigRcor}
\end{equation}
where
\begin{equation}
R^2 = { \widetilde{R}^2 - \mu^2 \tau b/a
	\over 1 + \bar\varepsilon b/a }
\label{eq:Rtrue}
\end{equation}
is the ``true'' form factor ratio, corrected for 2$\gamma$ exchange
effects, and $\widetilde{R}$ is the ``effective'' ratio, contaminated by
2$\gamma$ exchange. Note that in Eqs.~(\ref{eq:sigRcor}) and
(\ref{eq:Rtrue}) we have effectively linearized the quadratic term in
$\varepsilon$ by taking the average value of $\varepsilon$ ({\em i.e.},
$\bar\varepsilon$) over the $\varepsilon$ range being fitted. In
contrast to Ref.~\cite{Blu03}, where the approximation $a \approx 1$ was
made and the quadratic term in $\varepsilon$ neglected, the use of the
full expression in Eq.~(\ref{eq:Rtrue}) leads to a small decrease in $R$
compared with the approximate form.

\begin{figure}
\includegraphics[height=12cm,angle=270]{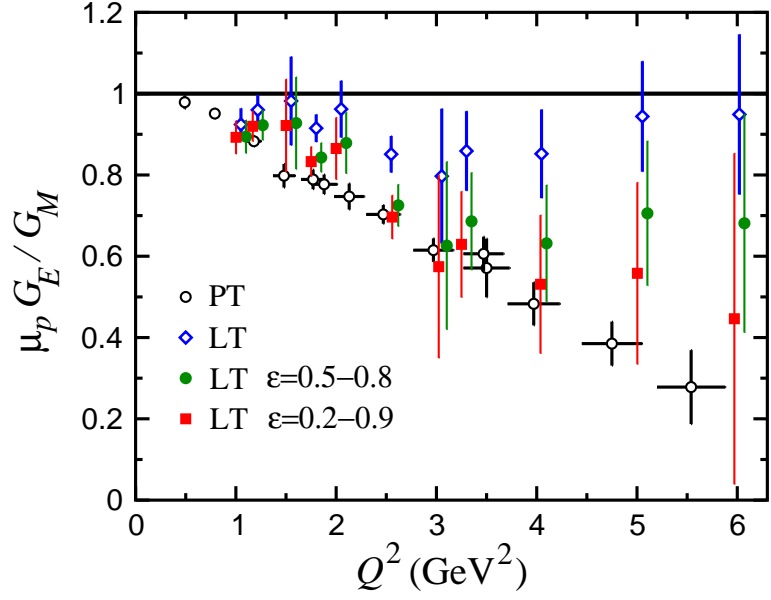}
\caption{The ratio of proton form factors $\mu_p G_E/G_M$
	measured using LT separation (open diamonds) \cite{Arr03}
	and polarization transfer (PT) (open circles) \cite{Jon00}.
	The LT points corrected for 2$\gamma$ exchange are shown
	assuming a linear slope for $\varepsilon=0.2-0.9$ (filled
	squares) and $\varepsilon=0.5-0.8$ (filled circles)
	(offset for clarity).
\label{fig:GEMp}}
\end{figure}

The shift in $R$ is shown in Fig.~\ref{fig:GEMp}, together with the
polarization transfer data. We consider two ranges for $\varepsilon$: a
large range $\varepsilon=0.2-0.9$, and a more restricted range
$\varepsilon=0.5-0.8$. The approximation of linear $\varepsilon$
dependence of $\Delta$ should be better for the latter, even though in
practice experiments typically sample values of $\varepsilon$ near its
lower and upper bounds. A proposed experiment at Jefferson Lab
\cite{Lin04} aims to test the linearity of the $\varepsilon$ plot
through a precision measurement of the unpolarized elastic cross
section.

The effect of the 2$\gamma$ exchange terms on $R$ is clearly
significant. As observed in Ref.~\cite{Blu03}, the 2$\gamma$ corrections
have the proper sign and magnitude to resolve a large part of the
discrepancy between the two experimental techniques. In particular, the
earlier results \cite{Blu03} using simple monopole form factors found a
shift similar to that in for the $\varepsilon=0.5-0.8$ range in
Fig.~\ref{fig:GEMp}, which resolves around 1/2 of the discrepancy. The
nonlinearity at small $\varepsilon$ makes the effective slope somewhat
larger if the $\varepsilon$ range is taken between 0.2 and 0.9. The
magnitude of the effect in this case is sufficient to bring the LT and
polarization transfer points almost to agreement, as indicated in
Fig.~\ref{fig:GEMp}.

While the 2$\gamma$ corrections clearly play a vital role in resolving
most of the form factor discrepancy, it is instructive to understand the
origin of the effect on $R$ with respect to contributions to the
individual $G_E^p$ and $G_M^p$ form factors. In general the amplitude
for elastic scattering of an electron from a proton, beyond the Born
approximation, can be described by three (complex) form factors,
$\widetilde F_1$, $\widetilde F_2$ and $\widetilde F_3$.
The generalized amplitude can be written as \cite{Gui03,Che04}
\begin{eqnarray}
{\cal M} &=& -i {e^2\over q^2} \ubar(p_3) \gamma_\mu u(p_1)\ 
\ubar(p_4) \left(\widetilde{F}_1\ \gamma^\mu\
 +\ \widetilde{F}_2\ {i \sigma^{\mu\nu} q_\nu \over 2 M}\
 +\ \widetilde{F}_3\ {\gamma\cdot K\ P^\mu \over M^2}\right) u(p_2)\ ,
\label{eq:current_gen}
\end{eqnarray}
where $K = (p_1 + p_3)/2$ and $P = (p_2 + p_4)/2$. The functions
$\widetilde{F}_i$ (both real and imaginary parts) are in general
functions of $Q^2$ and $\varepsilon$. In the 1$\gamma$ exchange limit
the $\widetilde{F}_{1,2}$ functions approach the usual (real) Dirac and
Pauli form factors, while the new form factor $\widetilde F_3$ exists
only at the 2$\gamma$ level and beyond,
\begin{eqnarray}
\widetilde{F}_{1,2}(Q^2,\varepsilon) &\to& F_{1,2}(Q^2)\ , \\
\widetilde{F}_3(Q^2,\varepsilon) &\to& 0\ .
\end{eqnarray}

Alternatively, the amplitude can be expressed in terms of the generalized
(complex) Sachs electric and magnetic form factors, $\widetilde{G}_E =
G_E + \delta G_E$ and $\widetilde{G}_M = G_M + \delta G_M$, in which case 
the reduced cross section, up to order $\alpha^2$ corrections, can be 
written \cite{Che04}
\begin{eqnarray}
\widetilde\sigma_R
&=& G_M^2 + {\varepsilon \over \tau} G_E^2
 +  2 G_M^2 \Re
    \left\{ {\delta G_M \over G_M} + \varepsilon Y_{2\gamma}
    \right\}
 + {2 \varepsilon \over \tau} G_E^2 \Re
    \left\{ {\delta G_E \over G_E} + {G_M \over G_E} Y_{2\gamma}
    \right\}\ ,
\end{eqnarray}
where the form factor $\widetilde{F}_3$ has been expressed in terms of
the ratio
\begin{eqnarray}
Y_{2\gamma}
&=& {\widetilde\nu {\widetilde{F}_3 \over G_M}} \ ,
\end{eqnarray}
with $\widetilde\nu \equiv K \cdot P/M^2
= \sqrt{\tau (1+\tau) (1+\varepsilon)/(1-\varepsilon)}$.
We should emphasize that the generalized form factors are not observables,
and therefore have no intrinsic physical meaning. Thus the magnitude and 
$\varepsilon$ dependence of the generalized form factors will depend on 
the choice of parametrization of the generalized amplitude.
For example, the axial parametrization introduces an effective axial vector
coupling beyond Born level, and is written as \cite{Rek}
\begin{eqnarray}
{\cal M} &=& -i {e^2\over q^2} \Biggl\{\ubar(p_3) \gamma_\mu u(p_1)\ 
\ubar(p_4) \left(F'_1\ \gamma^\mu\
 +\ F'_2\ {i \sigma^{\mu\nu} q_\nu \over 2 M}\right) u(p_2)\nonumber\\
&& +\ G'_A \ubar(p_3) \gamma_\mu \gamma_5 u(p_1)\ \ubar(p_4) 
\gamma^\mu\gamma_5 u(p_2)\Biggr\}\ .
\label{eq:current_gen2}
\end{eqnarray}
Following Ref.~\cite{Afa05}, one finds the relationships
\begin{eqnarray}
F'_1 &=& \widetilde{F}_1 + \widetilde{\nu}\widetilde{F}_3\ ,	\\
F'_2 &=& \widetilde{F}_2\ ,					\\
G'_A &=& -\tau \widetilde{F}_3\ .   \label{eq:genfftrans}
\end{eqnarray}

\begin{figure}[hbt]
\includegraphics[height=9cm,angle=270]{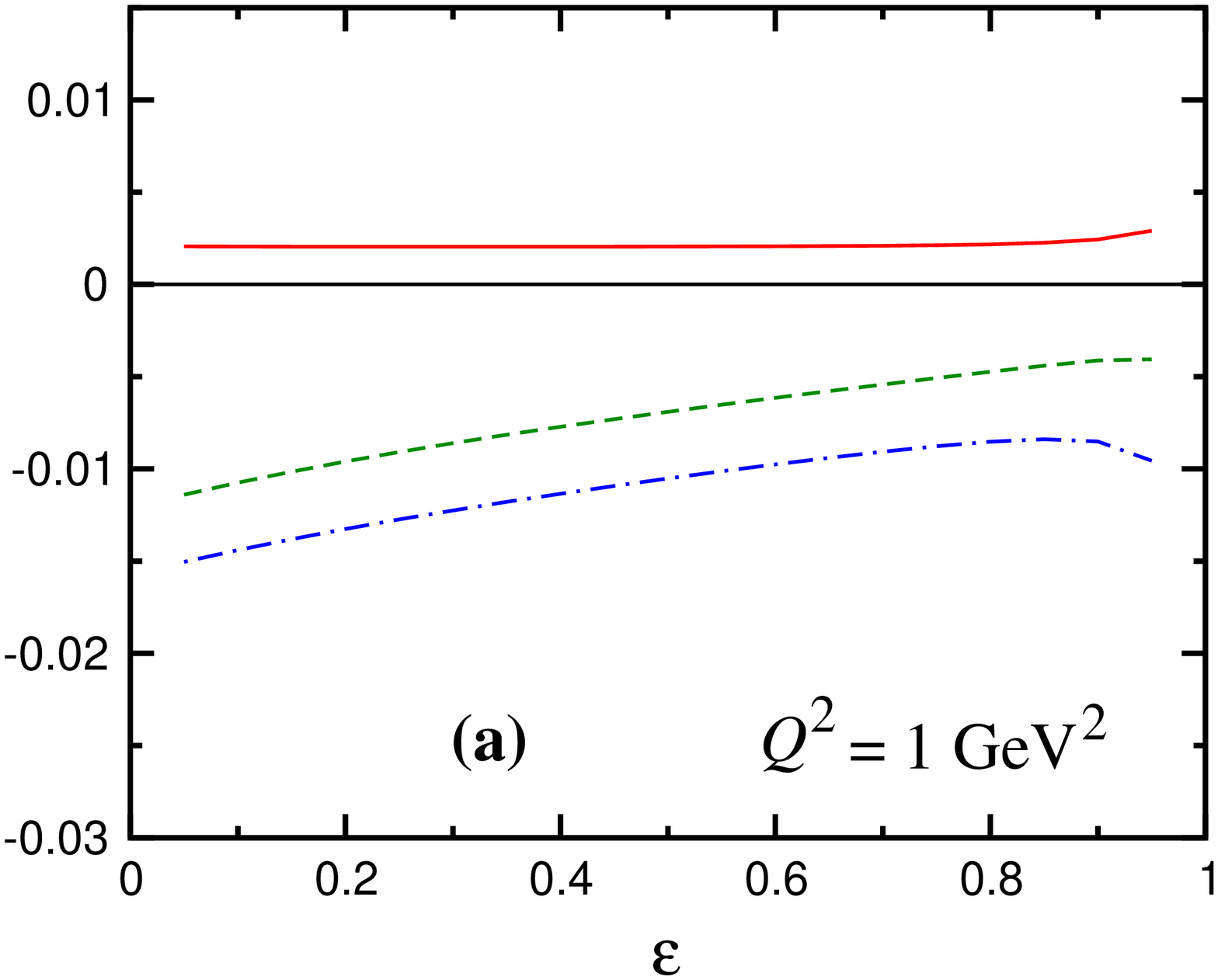}
\includegraphics[height=9cm,angle=270]{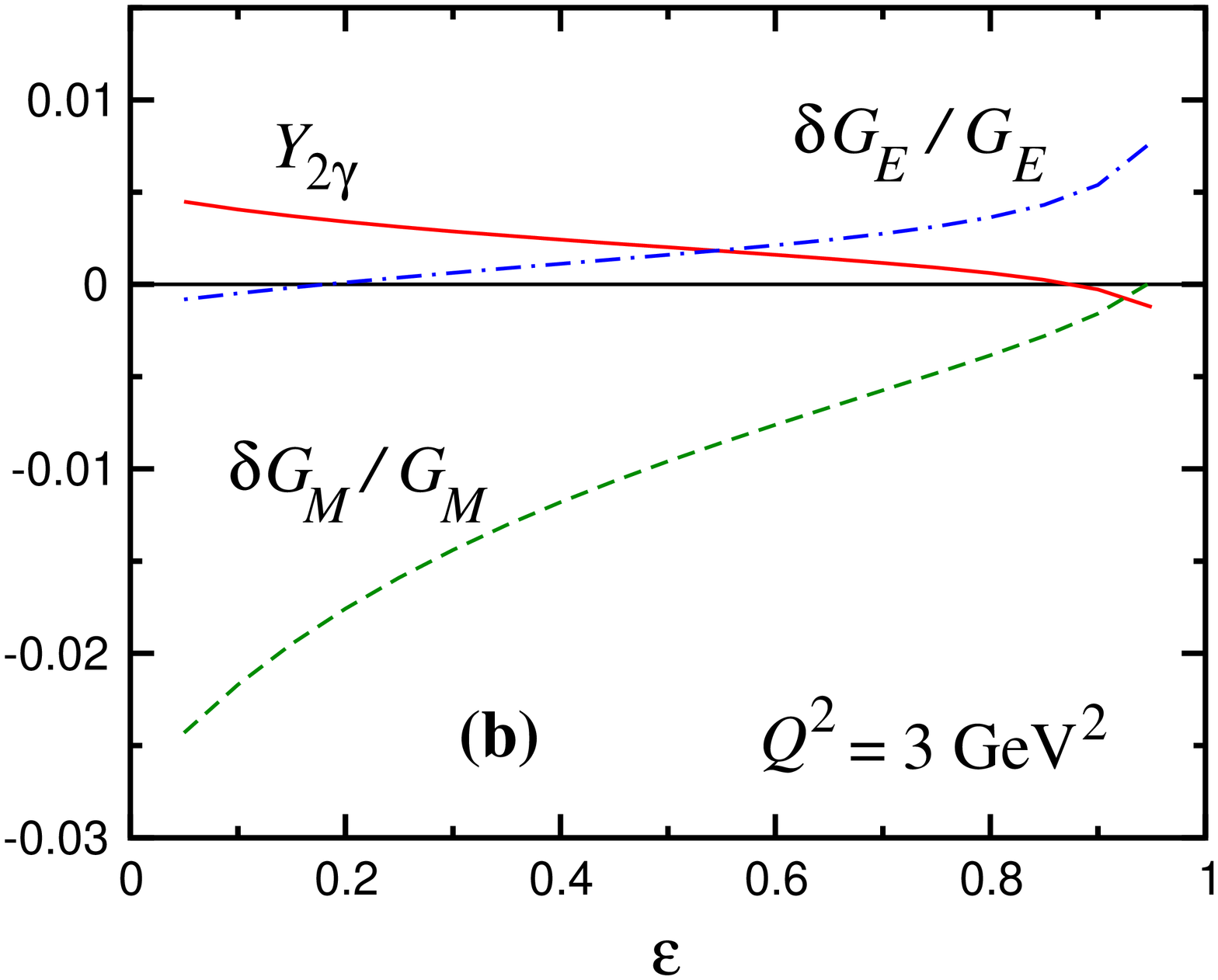}
\includegraphics[height=9cm,angle=270]{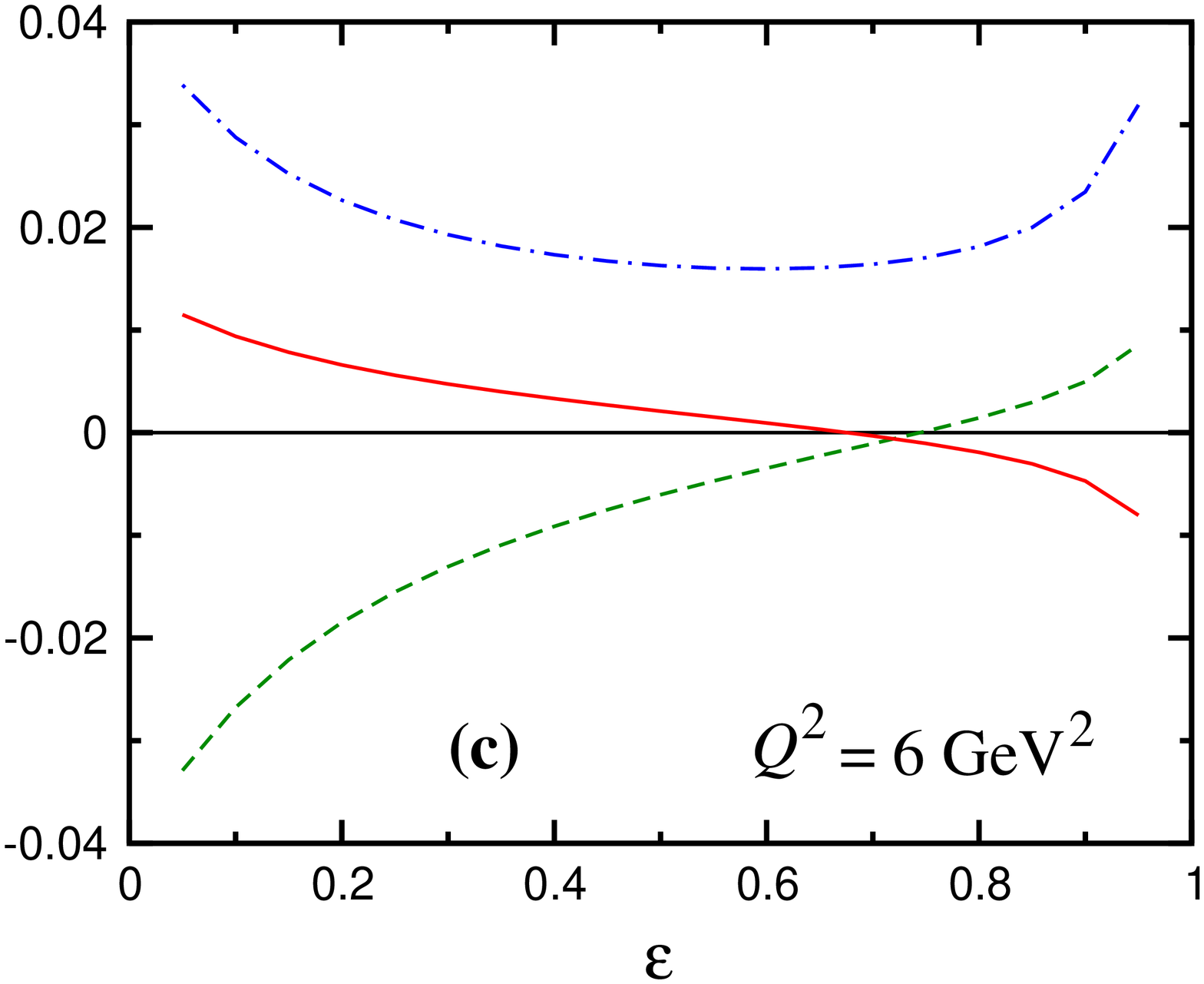}
\caption{Finite 2$\gamma$ contributions (defined with respect to the
	Mo-Tsai IR result \cite{MT69}) to the real parts of the $G_M$
	(dashed), $G_E$ (dot-dashed) and $Y_{2\gamma}$ (solid) form
	factors of the proton at $Q^2 = 1$, 3 and 6~GeV$^2$.
	Note the larger scale in the bottom figure.
\label{fig:df}}
\end{figure}

In Fig.~\ref{fig:df} we show the contributions of 2$\gamma$ exchange to
the (real parts of the) proton $\widetilde{G}_E$ and $\widetilde{G}_M$ 
form factors, and the
ratio $Y_{2\gamma}$ evaluated at $Q^2 = 1$, 3 and 6~GeV$^2$. One
observes that the 2$\gamma$ correction to $\widetilde{G}_M$ is large, with a
positive slope in $\varepsilon$ which increases with $Q^2$. The
correction to $\widetilde{G}_E$ is similar to that for $\widetilde{G}_M$ 
at $Q^2=1$~GeV$^2$, but
becomes shallower at intermediate $\varepsilon$ values for larger $Q^2$.
Both of these corrections are significantly larger than the
$Y_{2\gamma}$ correction, which is weakly $Q^2$ dependent, and has a
small negative slope in $\varepsilon$ at larger $Q^2$. The contribution
to $Y_{2\gamma}$ is found to be about 5 times smaller than that
extracted in phenomenological analyses \cite{Gui03} under the assumption
that the entire form factor discrepancy is due to the new $\widetilde{F}_3$
contribution (see also Ref.~\cite{Arr05}).

\subsection{Comparison of $e^+ p$ to $e^- p$ cross sections}

Direct experimental evidence for the contribution of 2$\gamma$
exchange can be obtained by comparing $e^+p$ and $e^-p$ cross
sections through the ratio
\begin{eqnarray}
R^{e^+e^-}
&\equiv& { d\sigma^{(e^+)} \over d\sigma^{(e^-)} }		\nonumber \\
&\approx& { \left| {\cal M}_0^{(e^+)} \right|^2
	  + 2 \Re \left\{ {\cal M}_0^{(e^+) \dagger}
			 {\cal M}^{2\gamma (e^+)}
		  \right\}
      \over \left| {\cal M}_0^{(e^-)} \right|^2
	  + 2 \Re \left\{ {\cal M}_0^{(e^-) \dagger}
			 {\cal M}^{2\gamma (e^-)}
		  \right\} }\ .
\end{eqnarray}
Whereas the Born amplitude ${\cal M}_0$ changes sign under the
interchange $e^- \leftrightarrow e^+$, the 2$\gamma$ exchange
amplitude ${\cal M}^{2\gamma}$ does not.
The interference of the ${\cal M}_0$ and ${\cal M}^{2\gamma}$
amplitudes therefore has the opposite sign for electron and positron
scattering.
Since the finite part of the 2$\gamma$ contribution is negative over
most of the range of $\varepsilon$, one would expect to see an
enhancement of the ratio of $e^+$ to $e^-$ cross sections,
\begin{eqnarray}
R^{e^+e^-}
&\approx& 1 - 2 \Delta\ , 
\end{eqnarray}
where $\Delta$ is defined in Eq.~(\ref{eq:Delta_dif}).

\begin{figure}
\includegraphics[height=12cm,angle=270]{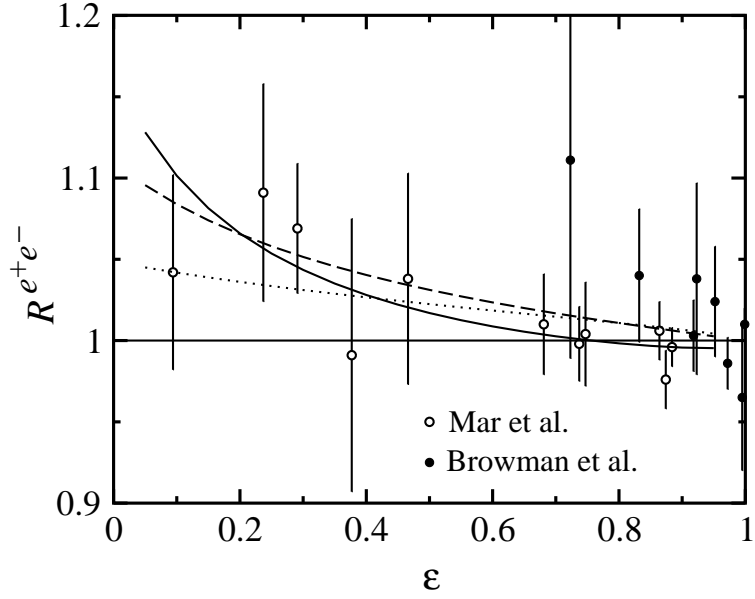}
\caption{Ratio of elastic $e^+ p$ to $e^-p$ cross sections.
	The data are from SLAC \cite{Bro65,Mar68}, with $Q^2$
	ranging from 0.01 to 5~GeV$^2$.
	The results of the 2$\gamma$ exchange calculations are
	shown by the curves for $Q^2 = 1$ (dotted), 3 (dashed)
	and 6~GeV$^2$ (solid).
\label{fig:Ree}}
\end{figure}

Although the current data on elastic $e^- p$ and $e^+ p$ scattering are
sparse, there are some experimental constraints from old data taken at
SLAC \cite{Bro65,Mar68}, Cornell \cite{And66}, DESY \cite{Bar67} and
Orsay \cite{Bou68} (see also Ref.~\cite{ArrEE}). The data are
predominantly at low $Q^2$ and at forward scattering angles,
corresponding to large $\varepsilon$ ($\varepsilon \agt 0.7$), where the
2$\gamma$ exchange contribution is small ($\alt 1\%$). Nevertheless, the
overall trend in the data reveals a small enhancement in $R^{e^+e^-}$ at
the lower $\varepsilon$ values, as illustrated in Fig.~\ref{fig:Ree}
(which shows a subset of the data, from the SLAC experiments
\cite{Bro65,Mar68}).

The data in Fig.~\ref{fig:Ree} are compared with our theoretical
results, calculated for several fixed values of $Q^2$ ($Q^2 = 1$, 3 and
6~GeV$^2$). The results are in good agreement with the data, although
the errors on the data points are quite large. Clearly better quality
data at backward angles, where an enhancement of up to $\sim 10\%$ is
predicted, would be needed for a more definitive test of the 2$\gamma$
exchange mechanism. An experiment \cite{Bro04} using a beam of $e^+ e^-$
pairs produced from a secondary photon beam at Jefferson Lab will make
simultaneous measurements of $e^- p$ and $e^+ p$ elastic cross sections
up to $Q^2 \sim 2$~GeV$^2$. A proposal to perform a precise ($\sim 1\%$)
comparison of $e^- p$ and $e^+ p$ scattering at $Q^2=1.6$~GeV$^2$ and
$\varepsilon \approx 0.4$ has also been made at the VEPP-3 storage ring
\cite{VEPP}.

\section{Polarized electron--proton scattering}

The results of the 2$\gamma$ exchange calculation in the previous
section give a clear indication of a sizable correction to the
LT-separated data at moderate and large $Q^2$. The obvious question
which arises is whether, and to what extent, the 2$\gamma$ exchange
affects the polarization transfer results, which show the dramatic
fall-off of the $G_E^p/G_M^p$ ratio at large $Q^2$. In this section we
examine this problem in detail.

The polarization transfer experiment involves the scattering of
longitudinally polarized electrons from an unpolarized proton target,
with the detection of the polarization of the recoil proton, $\vec e + p
\to e + \vec p$. (The analogous process whereby a polarized electron
scatters elastically from a polarized proton leaving an unpolarized
final state gives rise to essentially the same information.)
In the Born approximation the spin dependent amplitude is given by
\begin{eqnarray}
{\cal M}_0(s_1,s_4)
&=& -i {e^2 \over q^2}
\ubar(p_3) \gamma_\mu u(p_1,s_1)\
\ubar(p_4,s_4) \Gamma^\mu (q) u(p_2)\ ,
\end{eqnarray}
where $s_1 = (s_1^0; \vec s_1)$ and $s_4 = (s_4^0; \vec s_4)$
are the spin four-vectors of the initial electron and final proton,
respectively, and the spinor $u(p_1,s_1)$ is defined such that
$u(p_1,s_1) \ubar(p_1,s_1)
= (\slash{p}_1+m)(1+\gamma_5 \slash{s}_1)/2$, and similarly for
$\ubar(p_4,s_4)$.
The spin four-vector (for either the electron or recoil proton) can be
written in terms of the 3-dimensional spin vector $\zeta$ specifying
the spin direction in the rest frame
(see {\em e.g.} Ref.~\cite{MP00}),
\begin{eqnarray}
s^\mu &=&
\left( { \vec\zeta \cdot \vec p \over m };
       \vec\zeta + \vec p { \vec\zeta \cdot \vec p \over m (m+E) }
\right)\ ,
\label{eq:s}
\end{eqnarray}
where $m$ and $E$ are the mass and energy of the electron or proton.
Clearly in the limit $\vec p \to 0$, the spin four-vector
$s \to (0; \vec\zeta)$.
Since $\zeta$ is a unit vector, one has $\vec\zeta^2 = 1$,
and one can verify from Eq.~(\ref{eq:s}) that $s^2 = -1$
and $p \cdot s = 0$.
If the incident electron energy $E_1$ is much larger than the electron
mass $m$, the electron spin four-vector can be related to the electron
helicity $h = \vec\zeta_1 \cdot \vec p_1$ by
\begin{eqnarray}
s_1 &\approx& h\ { p_1 \over m }\ .
\label{eq:s1}
\end{eqnarray}

The coordinate axes are chosen so that the recoil proton momentum
$\vec p_4$ defines the $z$ axis, in which case for longitudinally
polarized protons one has $\vec\zeta = \hat p_4$.
In the 1$\gamma$ exchange approximation the elastic cross section
for scattering a longitudinally polarized electron with a recoil
proton polarized longitudinally is then given by
\begin{eqnarray}
{ d\sigma^{(L)} \over d\Omega }
&=& h\ \sigma_{\rm Mott}\
   {E_1 + E_3 \over M} \sqrt{\tau \over 1+\tau} \tan^2{\theta\over 2}\
    G_M^2\ .
\end{eqnarray}

For transverse recoil proton polarization we define the $x$ axis
to be in the scattering plane, $\hat x = \hat y\ \times \hat z$,
where $\hat y = \hat p_1 \times \hat p_3$ defines the direction
perpendicular, or normal, to the scattering plane.
The elastic cross section for producing a transversely polarized
proton in the final state, with $\vec\zeta \cdot \vec p_4 = 0$,
is given by
\begin{eqnarray}
{ d\sigma^{(T)} \over d\Omega }
&=& h\ \sigma_{\rm Mott}\
    2 \sqrt{\tau \over 1+\tau} \tan{\theta\over 2}\ 
    G_E\ G_M\ .
\end{eqnarray}
Taking the ratio of the transverse to longitudinal proton cross
sections then gives the ratio of the electric to magnetic proton
form factors, as in Eq.~(\ref{eq:poltrans}).
Note that in the 1$\gamma$ exchange approximation the normal
polarization is identically zero.

The amplitude for the 2$\gamma$ exchange diagrams in
Fig.~\ref{fig:diag} with the initial electron and final proton
polarized can be written as
\begin{eqnarray}
{\cal M}^{2\gamma}(s_1,s_4)
&=& e^4 \int {d^4 k\over (2\pi)^4}
    {N_{\rm box}(k,s_1,s_4) \over D_{\rm box}(k)}
	+ e^4 \int {d^4 k\over (2\pi)^4}{N_{\rm x-box}(k,s_1,s_4) \over D_{\rm x-box}(k)}
    \ ,
\label{eq:mbox_pol}
\end{eqnarray}
where the numerators are the matrix elements
\begin{eqnarray}
N_{\rm box}(k,s_1,s_4)
&=& \ubar(p_3) \gamma_\mu (\slash{p}_1 - \slash{k} + m)
    \gamma_\nu u(p_1,s_1)				\nonumber\\
&\times& \ubar(p_4,s_4) \Gamma^\mu(q-k) (\slash{p}_2 + \slash{k} + M)
	 \Gamma^\nu(k) u(p_2)\ ,
\label{eq:nbox_pol}					\\
N_{\rm x-box}(k,s_1,s_4)
&=& \ubar(p_3) \gamma_\nu (\slash{p}_3 + \slash{k} + m)
    \gamma_\mu u(p_1,s_1)				\nonumber\\
&\times& \ubar(p_4,s_4) \Gamma^\mu(q-k) (\slash{p}_2 + \slash{k} + M)
	 \Gamma^\nu(k) u(p_2)\ ,
\label{eq:nxbox_pol}
\end{eqnarray}
and the denominators are given in Eqs.~(\ref{eq:dbox}) and
(\ref{eq:dxbox}).
The traces in Eqs.~(\ref{eq:nbox_pol}) and (\ref{eq:nxbox_pol})
can be evaluated using the explicit expression for the spin-vectors
$s_1$ and $s_4$ in Eqs.~(\ref{eq:s}) and (\ref{eq:s1}).

In analogy with the unpolarized case (see Eq.~(\ref{eq:Delta_dif})),
the spin-dependent corrections to the longitudinal ($\Delta_L$) and
transverse ($\Delta_T$) cross sections are defined as the
finite parts of the 2$\gamma$ contributions relative to the IR
expression from Mo \& Tsai \cite{MT69} in Eq.~(\ref{eq:deltaIRMT}),
which are independent of polarization,
\begin{eqnarray}
\Delta_{L,T} &=& \delta_{L,T}^{\rm full} - \delta_{\rm IR}\ .
\end{eqnarray}
Experimentally, one does not usually measure the longitudinal or
transverse cross section {\em per se}, but rather the 
ratio of the transverse or longitudinal cross section to the unpolarized
cross section, denoted $P_L$ or $P_T$, respectively. Thus the 2$\gamma$
exchange correction to the polarization transfer ratio can be incorporated as
\begin{equation}
{P_{L,T}^{1\gamma+2\gamma}\over P_{L,T}^{1\gamma}} = {1+\Delta_{L,T}\over
1+\Delta}\ ,
\label{eq:polcorr}
\end{equation}
where $\Delta$ is the correction to the unpolarized cross section
considered in the previous section.

\begin{figure}
\includegraphics[height=12cm,angle=270]{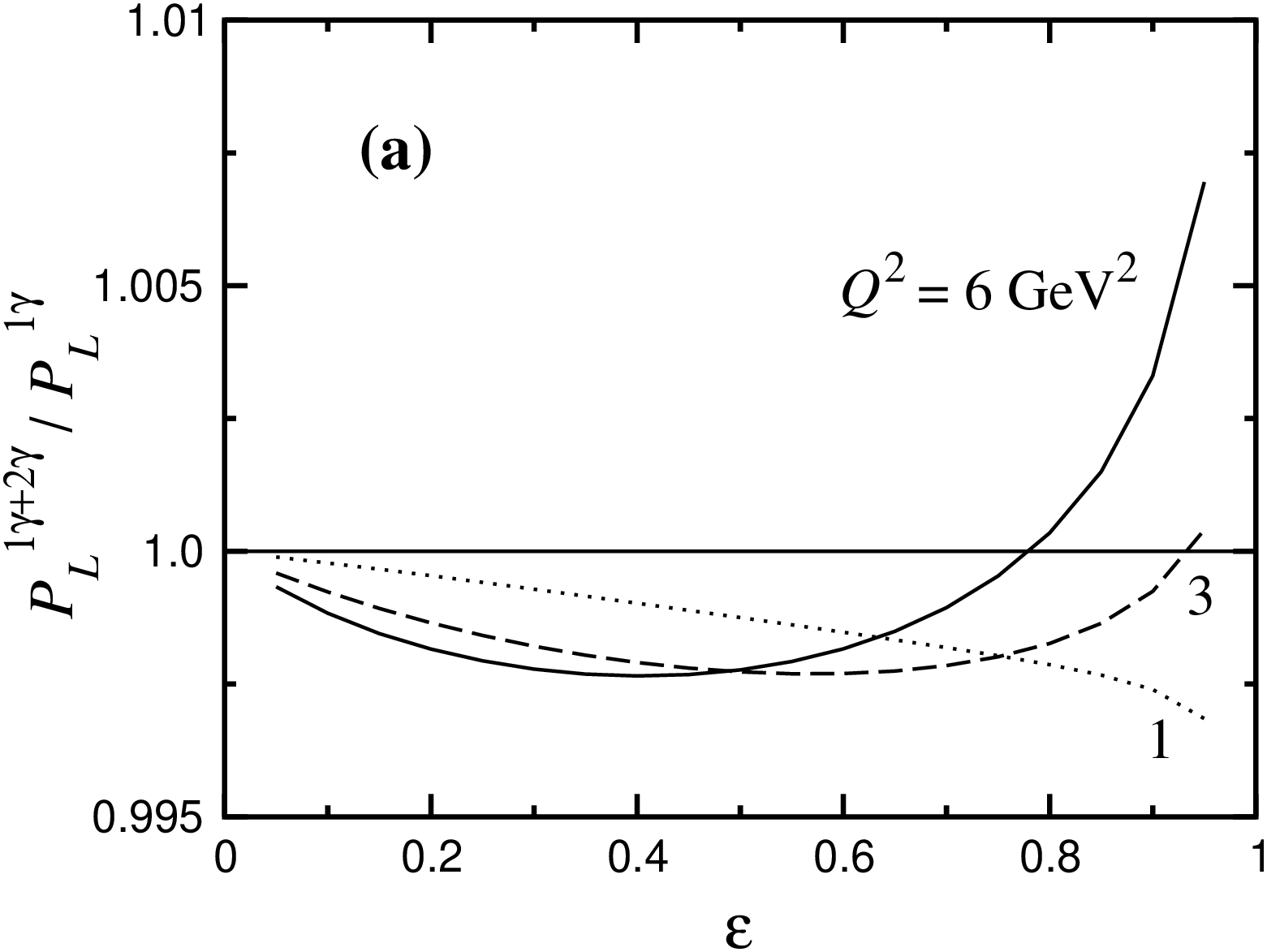}\\
\vspace*{-1.25cm}
\includegraphics[height=12cm,angle=270]{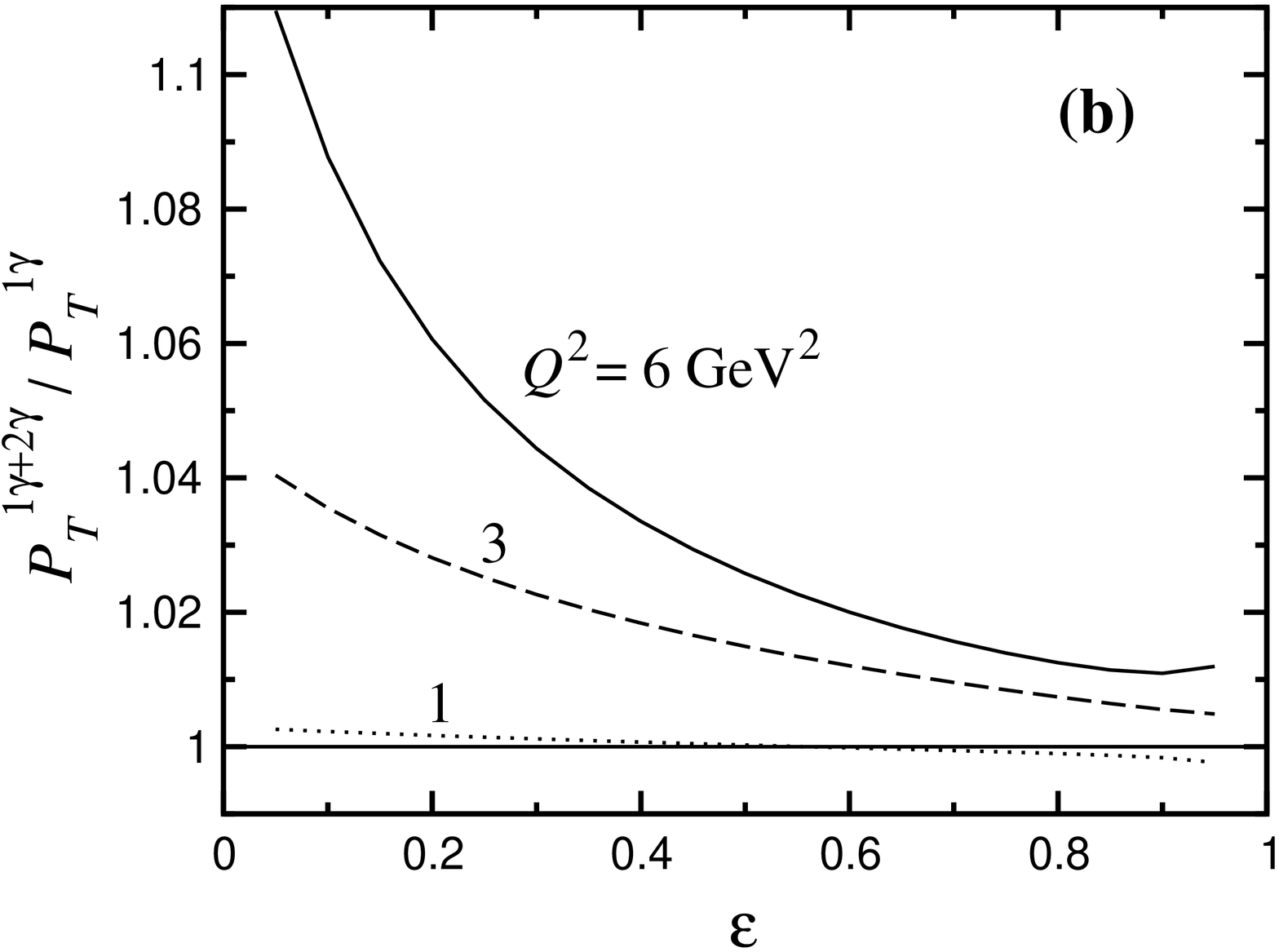}
\caption{Ratio of the finite part (with respect to the IR contribution
	in Eq.~(\ref{eq:deltaIR})) of the Born+2$\gamma$ correction relative
	to the Born term, for (a) longitudinal and (b) transverse
	recoil proton polarization, at $Q^2 = 1$ (dotted), 3 (dashed)
	and 6~GeV$^2$ (solid).  Note the different scales on the vertical
	axes.
\label{fig:delLT}}
\end{figure}

The 2$\gamma$ exchange contribution relative to the Born term is shown
in Fig.~\ref{fig:delLT}. The correction to the longitudinal polarization
transfer ratio $P_L$ is small overall. This is because the correction
$\Delta_L$ to the longitudinal cross section is roughly the same as the
correction $\Delta$ to the unpolarized cross section. The corrections
$\Delta$ and $\Delta_L$ must be exactly the same at $\theta=180^\circ$ 
($\varepsilon=0$), and our
numerical results bear this out. By contrast, the correction to the
transverse polarization transfer ratio $P_T$ is enhanced at backward
angles, and grows with $Q^2$. This is due to a combined effect of
$\Delta_T$ becoming more positive with increasing $Q^2$, and $\Delta$
becoming more negative.

\begin{figure}
\includegraphics[height=12cm,angle=270]{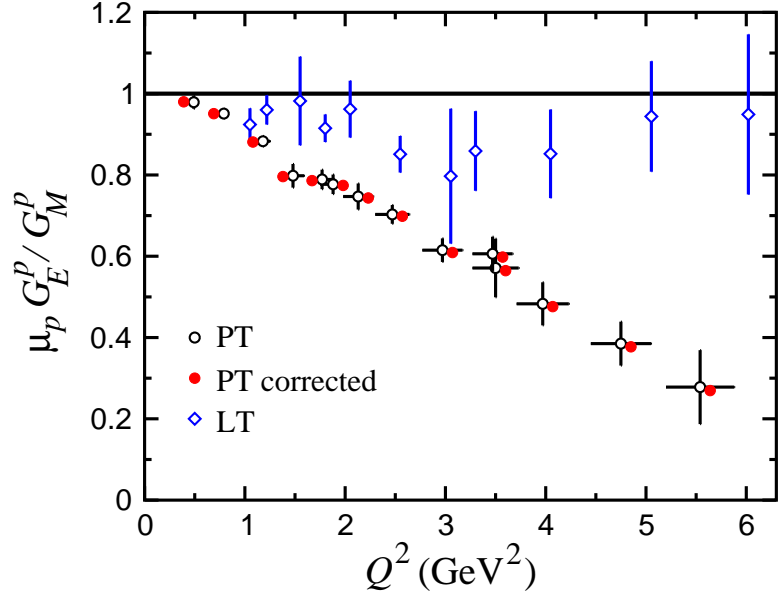}
\caption{Proton electric to magnetic form factor ratio obtained from
	the polarization transfer measurements \protect\cite{Jon00},
	with (solid circles) and without (open circles) the 2$\gamma$
	exchange corrections.  The corrected values have been offset
	for clarity.
	The LT-separated ratio (open diamonds) from
	Fig.~\protect\ref{fig:GEMp} is shown for comparison.
\label{fig:GEMpt}}
\end{figure}

In the standard radiative corrections using the results of Mo \& Tsai
\cite{MT69}, the corrections for transverse polarization are the same
as those for longitudinal polarization, so that no additional
corrections beyond hard bremsstrahlung need be applied \cite{MP00}.
Because the polarization transfer experiments \cite{Jon00} typically
have $\varepsilon \approx 0.7$--0.8, the shift in the polarization
transfer ratio in Eq.~(\ref{eq:poltrans}) due to the 2$\gamma$
exchange corrections is not expected to be dramatic.
If $R$ is the corrected (``true'') electric to magnetic form factor
ratio, as in Eq.~(\ref{eq:sigRcor}), then the measured polarization
transfer ratio is
\begin{eqnarray}
\widetilde R &=& R
  \left( {1 + \Delta_T \over 1 + \Delta_L}
  \right)\ .
\label{eq:shiftedR}
\end{eqnarray}
Inverting Eq.~(\ref{eq:shiftedR}), the shift in the ratio $R$ is
illustrated in Fig.~\ref{fig:GEMpt} by the filled circles (offset
slightly for clarity).
The unshifted results are indicated by the open circles, and the
LT separated results are labeled by diamonds.
The effect of the 2$\gamma$ exchange on the form factor ratio is a very 
small, $\alt 3\%$ suppression of the ratio at the larger $Q^2$ values, 
which is well within the experimental uncertainties.

Note that the shift in $R$ in Eq.~(\ref{eq:shiftedR}) does not include
corrections due to hard photon bremsstrahlung (which are part of the
standard radiative corrections).
Since these would make both the numerator and denominator in
Eq.~(\ref{eq:shiftedR}) even larger, the correction shown in
Fig.~\ref{fig:GEMpt} would represent an upper limit on the shift in $R$.

\begin{figure}
\includegraphics[height=12cm,angle=270]{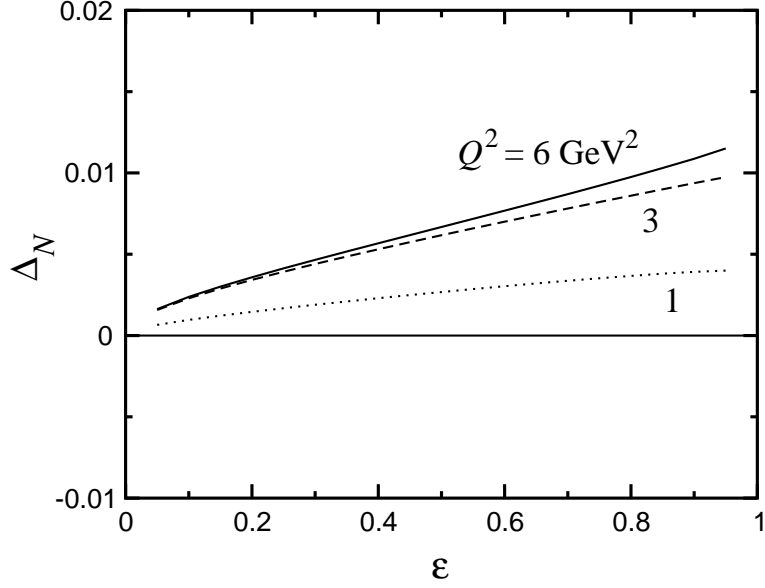}
\caption{Ratio of the 2$\gamma$ contribution to the normal polarization,
	to the unpolarized Born contribution,
	as a function of $\varepsilon$, for $Q^2 = 1$ (dotted),
        3 (dashed) and 6~GeV$^2$ (solid).
\label{fig:delN}}
\end{figure}

Finally, the 2$\gamma$ exchange process can give rise to a non-zero
contribution to the elastic cross section for a recoil proton polarized
normal to the scattering plane. This contribution is purely imaginary,
and does not exist in the 1$\gamma$ exchange approximation. It is
illustrated in Fig.~\ref{fig:delN}, where the ratio $\Delta_N$ of the
2$\gamma$ exchange contribution relative to the {\em unpolarized} Born
contribution is shown as a function of $\varepsilon$ for several values
of $Q^2$. (For consistency in notation we denote this correction
$\Delta_N$ rather than $\delta_N$, even though there is no IR 
contribution to the normal polarization.)

\begin{figure}
\includegraphics[height=12cm,angle=270]{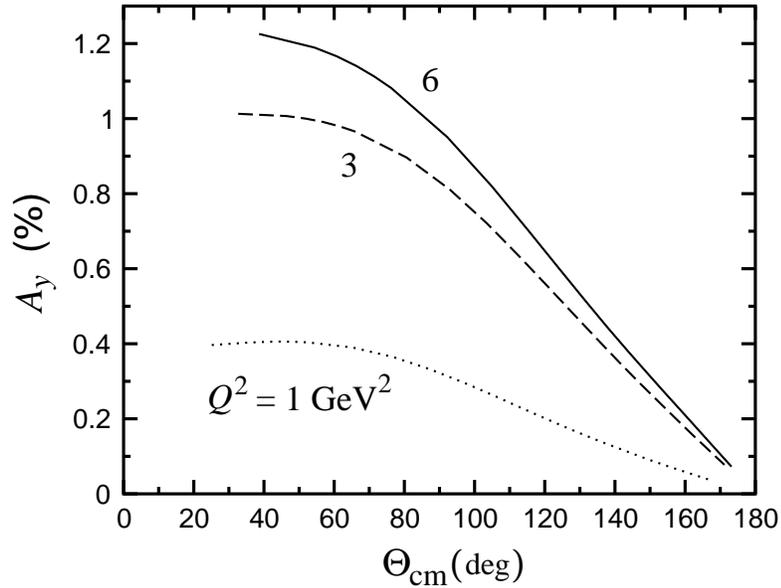}
\caption{Normal polarization asymmetry, expressed as a percentage,
	as a function of the center of mass scattering angle,
	$\Theta_{\rm cm}$, for $Q^2=1$ (dotted), 3 (dashed) and
	6~GeV$^2$ (solid).
\label{fig:Ay}}
\end{figure}

The normal polarization contribution is very small numerically,
$\Delta_N \alt 1\%$, and has a very weak $\varepsilon$ dependence. In
contrast to $\Delta_L$ and $\Delta_T$, the normal polarization ratio is
smallest at low $\varepsilon$, becoming larger with increasing
$\varepsilon$. Although not directly relevant to the elastic form factor
extraction, the observation of protons with normal polarization would
provide direct evidence of 2$\gamma$ exchange in elastic scattering.
Figure~\ref{fig:Ay} shows the normal polarization asymmetry $A_y$ as a
function of the center of mass scattering angle, $\Theta_{\rm cm}$, for
several values of $Q^2$. The asymmetry is relatively small, of the order
of 1\% at small $\Theta_{\rm cm}$ for $Q^2 \sim 3$~GeV$^2$, but grows
with $Q^2$.

The imaginary part of the 2$\gamma$ amplitude can also be accessed by
measuring the electron beam asymmetry for electrons polarized normal to
the scattering plane \cite{Wells}. Knowledge of the imaginary part of
the 2$\gamma$ exchange amplitude could be used to constrain models of
Compton scattering, although relating this to the real part (as needed
for form factor studies) would require a dispersion relation analysis.

\section{Electron--neutron scattering}

In this section we examine the effect of the 2$\gamma$ exchange
contribution on the form factors of the neutron. Since the magnitude of
the electric form factor of the neutron is relatively small compared
with that of the proton, and as we saw in Sec.~III the effects on the
proton are significant at large $Q^2$, it is important to investigate
the extent to which $G_E^n$ may be contaminated by 2$\gamma$ exchange.

\begin{figure}
\includegraphics[height=12cm,angle=270]{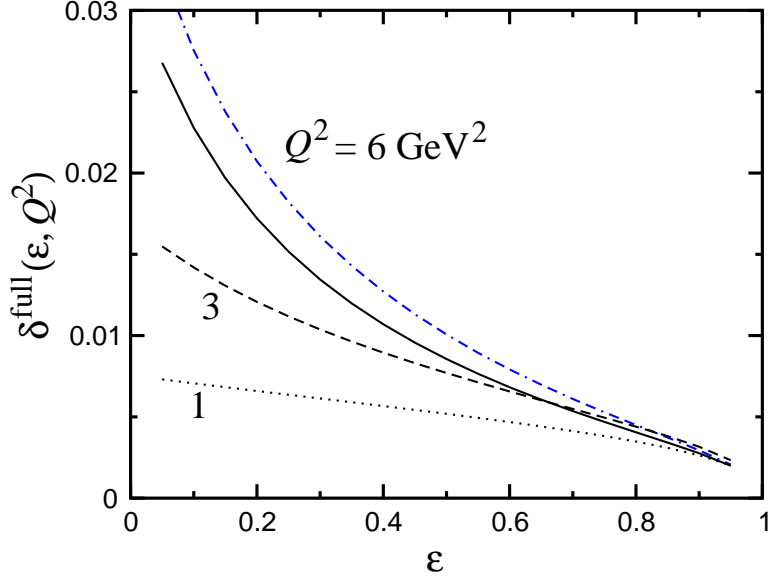}
\caption{2$\gamma$ contribution to the unpolarized electron--neutron
        elastic scattering cross section, at $Q^2=1$ (dotted),
	3 (dashed) and 6~GeV$^2$ (solid and dot-dashed).
	The dot-dashed curve corresponds to the form factor
	parameterization of Ref.~\cite{Bos95}, while the others are from
	Ref.~\cite{Mer96} (as fitted by the parameters in Table~I).
\label{fig:del_n}}
\end{figure}

Using the same formalism as in Secs.~II and III, the calculated
2$\gamma$ exchange correction for the neutron is shown in
Fig.~\ref{fig:del_n} for $Q^2 = 1$, 3 and 6~GeV$^2$.
Since there is no IR divergent contribution to $\delta$ for the
neutron, the total 2$\gamma$ correction $\delta^{\rm full}$ is
displayed in Fig.~\ref{fig:del_n}.
In the numerical calculation, the input neutron form factors from
Ref.~\cite{Mer96} are parameterized using the pole fit in
Eq.~(\ref{eq:3pole}), with the parameters given in
Table~\ref{tab:param}. For comparison, the correction at $Q^2=6$~GeV$^2$
is also computed using a 3-pole fit to the form factor parameterization
from Ref.~\cite{Bos95}. The difference between these is an indication of
the model dependence of the calculation.

The most notable difference with respect to the proton results is the
sign and slope of the 2$\gamma$ exchange correction. Namely, the
magnitude of the correction $\delta^{\rm full}(\varepsilon,Q^2)$ for the
neutron is $\sim 3$ times smaller than for the proton. The reason for
the sign change is the negative anomalous magnetic moment of the
neutron. The $\varepsilon$ dependence is approximately linear at
moderate and high $\varepsilon$, but at low $\varepsilon$ there exists a
clear deviation from linearity, especially at large $Q^2$.

\begin{figure}
\includegraphics[height=12cm,angle=270]{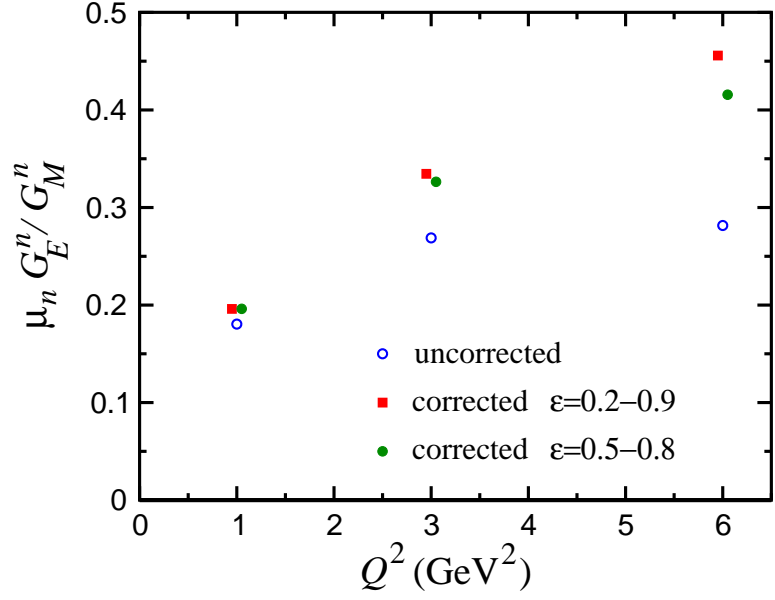}
\caption{Effect of 2$\gamma$ exchange on the ratio of neutron form
	factors $\mu_n G_E^n/G_M^n$ using LT separation.
	The uncorrected points (open circles) are from the form
	factor parameterization in Ref.~\protect\cite{Mer96},
	while the points corrected for 2$\gamma$ exchange are
	obtained from linear fits to $\delta^{\rm full}$ in
	Fig.~\protect\ref{fig:del_n} for $\varepsilon=0.2-0.9$
	(filled squares) and $\varepsilon=0.5-0.8$ (filled circles)
	(offset for clarity).
\label{fig:GEMn_LT}}
\end{figure}

Translating the $\varepsilon$ dependence to the form factor ratio, the
resulting shift in $\mu_n G_E^n/G_M^n$ is shown in
Fig.~\ref{fig:GEMn_LT} at several values of $Q^2$, assuming a linear
2$\gamma$ correction over two different $\varepsilon$ ranges
($\varepsilon=0.2-0.9$ and $\varepsilon=0.5-0.8$). The baseline
(uncorrected) data are from the global fit in Ref.~\cite{Mer96}. The
shift due to 2$\gamma$ exchange is small at $Q^2 = 1$~GeV$^2$, but
increases significantly by $Q^2 = 6$~GeV$^2$, where it produces a
50--60\% rise in the uncorrected ratio. These results suggest that, as
for the proton, the LT separation method is subject to large corrections
from 2$\gamma$ exchange at large $Q^2$.

\begin{figure}
\includegraphics[height=12cm,angle=270]{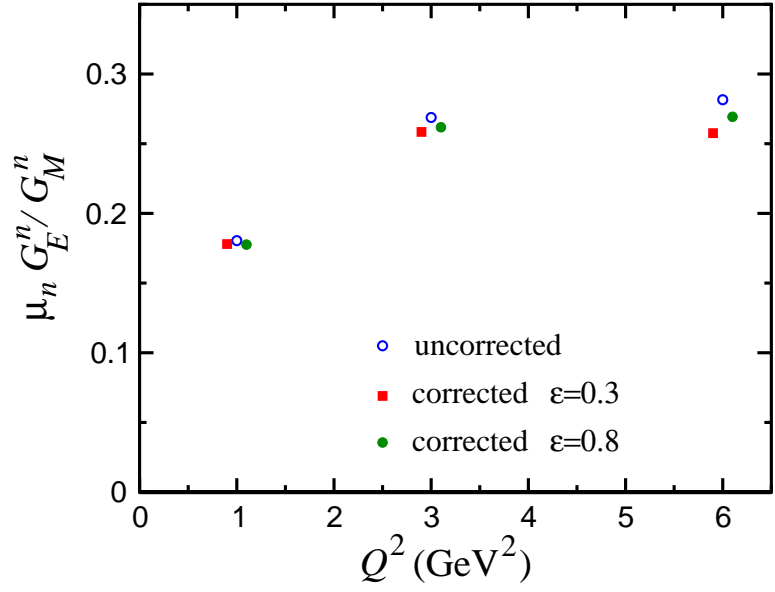}
\caption{Effect of 2$\gamma$ exchange on the ratio of neutron form
	factors $\mu_n G_E^n/G_M^n$ using polarization transfer.
	The uncorrected points (open circles) are from the
	parameterization in Ref.~\protect\cite{Mer96}, and
	the points corrected for 2$\gamma$ exchange correspond
	to $\varepsilon=0.3$ (filled squares) and $\varepsilon=0.8$
	(filled circles) (offset for clarity).
\label{fig:GEMn_PT}}
\end{figure}

While the 2$\gamma$ corrections to the form factor ratio from LT
separation are signficant, particularly at large $Q^2$, in practice the
neutron $G_E^n$ form factor is commonly extracted using the polarization
transfer method. To compare the 2$\gamma$ effects on the ratio $\mu_n
G_E^n/G_M^n$ extracted by polarization transfer, in
Fig.~\ref{fig:GEMn_PT} we plot the same ``data points'' as in
Fig.~\ref{fig:GEMn_LT}, shifted by the $\delta_{L,T}$ corrections as in
Eq.~(\ref{eq:shiftedR}) at two values of $\varepsilon$
($\varepsilon=0.3$ and 0.8). The shift is considerably smaller than
that from the LT method, but nevertheless represents an
approximately 4\% (3\%) suppression at $\varepsilon=0.3$ (0.8) for
$Q^2=3$~GeV$^2$, and $\approx 10\%$ (5\%) suppression for
$Q^2=6$~GeV$^2$ for the same $\varepsilon$. In the Jefferson Lab
experiment \cite{Madey} to measure $G_E^n/G_M^n$ at $Q^2=1.45$~GeV$^2$
the value of $\varepsilon$ was around 0.9, at which the 2$\gamma$
correction was $\approx 2.5\%$. In the recently approved extension of
this measurement to $Q^2 \approx 4.3$~GeV$^2$ \cite{MadeyNew}, the
2$\gamma$ correction for $\varepsilon \approx 0.82$ is expected to be
around 3\%. While small, these corrections will be important to take
into account in order to achieve precision at the several percent level.

\section{$^3$He Elastic Form Factors}

In this section we extend our formalism to the case of elastic
scattering from $^3$He nuclei. Of course, the contribution of $^3$He
intermediate states in 2$\gamma$ exchange is likely to constitute only a
part of the entire effect -- contributions from break-up channels may
also be important. However, we can obtain an estimate on the size of the
effect on the $^3$He form factors, in comparison with the effect on the
nucleon form factor ratio.

The expressions used to evaluate the 2$\gamma$ contributions are similar
to those for the nucleon, since $^3$He is a spin-${1 \over 2}$ particle,
although there are some important differences. For instance, the charge
is now $Z e$ (where $Z=2$ is the atomic number of $^3$He), the mass
$M_{^3{\rm He}}$ is $\approx 3$ times larger than the nucleon mass, and
the anomalous magnetic moment is $\kappa_{^3{\rm He}} = -4.185$. In
addition, the internal $\gamma {}^3{\rm He}$ form factor is somewhat
softer than the corresponding nucleon form factor (since the charge
radius of the $^3$He nucleus is $\approx 1.88$~fm). Using a dipole shape
for the form factor gives a cut-off mass of $\Lambda_{^3{\rm He}}
\approx 0.37$~GeV.

\begin{figure}
\includegraphics[height=12cm,angle=270]{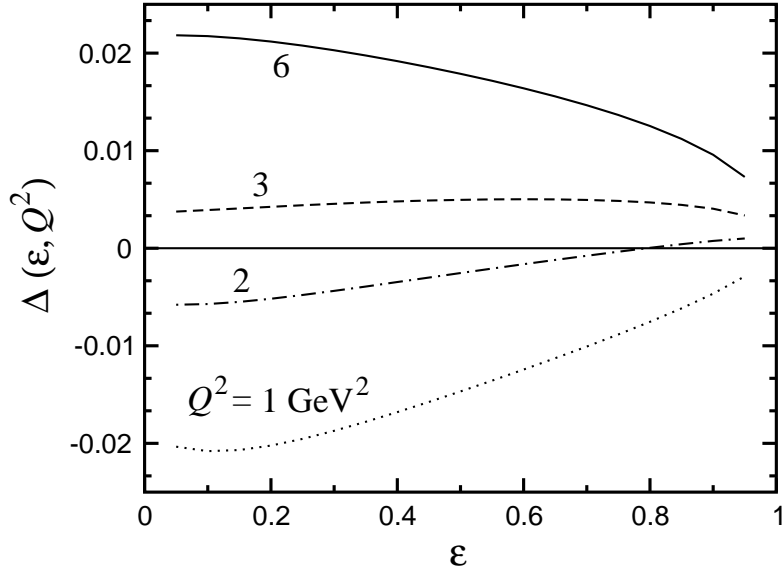}
\caption{2$\gamma$ contribution to the unpolarized electron--$^3$He
	elastic scattering cross section, with the $^3$He elastic
	intermediate state, as a function of $\varepsilon$, for
	$Q^2 = 1$ (dotted), 2 (dot-dashed), 3 (dashed) and 6~GeV$^2$
	(solid).
\label{fig:del_He3}}
\end{figure}

The 2$\gamma$ exchange correction is shown in Fig.~\ref{fig:del_He3} as
a function of $\varepsilon$ for several values of $Q^2$. The
$\varepsilon$ dependence illustrates the interesting interplay between
the Dirac and Pauli contributions to the cross section. At low $Q^2$
($Q^2 \sim 1$~GeV$^2$), the $F_1$ contribution is dominant, and the
effect has the same sign and similar magnitude as in the proton. The
result in fact reflects a partial cancellation of 2 opposing effects:
the larger charge squared $Z^2$ of the $^3$He nucleus makes the effect
larger (by a factor $\sim 4$), while the larger mass squared of the
$^3$He nucleus suppresses the effect by a factor $\sim 9$. In addition,
the form factor used is much softer than that of the nucleon, so that
the overall effect turns out to be similar in magnitude as for the
proton.

\begin{figure}
\includegraphics[height=12cm,angle=270]{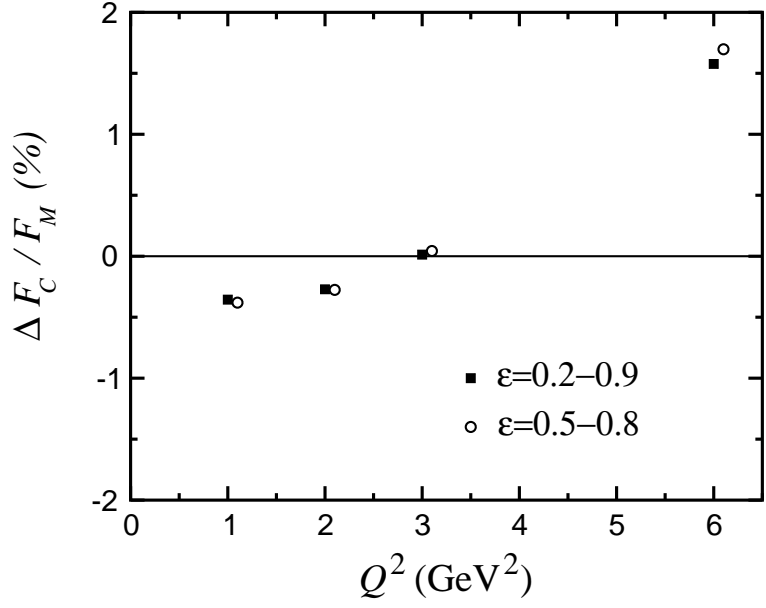}
\caption{Relative change (in percent) in the ratio of the charge to 
	magnetic form factors of $^3$He due to 2$\gamma$ exchange,
	obtained from linear fits to $\Delta$ in
	Fig.~\protect\ref{fig:del_He3} for $\varepsilon=0.2-0.9$
	(filled squares) and $\varepsilon=0.5-0.8$ (open circles)
	(offset for clarity).
\label{fig:R_He3}}
\end{figure}

With increasing $Q^2$ the Pauli $F_2$ term becomes more important, so
that for $Q^2 \agt 3$~GeV$^2$ the overall sign of the contribution is
positive. Interestingly, over most of the region between $\varepsilon
\approx 0.2$ and 0.9 the slope in $\varepsilon$ is approximately
constant. This allows us to extract the correction to the ratio of
charge to magnetic form factors, $F_C/F_M$, which we illustrate in
Fig.~\ref{fig:R_He3}. The effect is a small, $\alt 0.5\%$ reduction in
the ratio for $Q^2 \alt 3$~GeV$^2$, which turns into an enhancement at
large $Q^2$. However, the magnitude of the effect is small, and even for
$Q^2 = 6~$GeV$^2$ the 2$\gamma$ effect only gives $\alt 2\%$ increase in
the form factor ratio. Proposed experiments at Jefferson Lab \cite{Pet03} 
would measure the $^3$He form factors to $Q^2 \approx 4$~GeV$^2$.

\section{Conclusion}

We have presented a comprehensive analysis of the effects of 2$\gamma$
exchange in elastic electron--nucleon scattering, taking particular
account of the effects of nucleon structure. Our main purpose has been
to quantify the 2$\gamma$ effect on the ratio of electric to magnetic
form factors of the proton, which has generated controversy recently
stemming from conflicting results of measurements at large $Q^2$.

Consistent with the earlier preliminary investigation \cite{Blu03}, we
find that inclusion of 2$\gamma$ exchange reduces the $G_E^p/G_M^p$
ratio extracted from LT-separated cross section data, and resolves a
significant amount of the discrepancy with the polarization transfer
results. At higher $Q^2$ we find strong deviations from linearity,
especially at small $\varepsilon$, which can be tested in future
high-precision cross section measurements. There is some residual
model-dependence in the calculation of the 2$\gamma$ amplitude arising
from the choice of form factors at the internal $\gamma^* NN$ vertices
in the loop integration. This dependence, while not overwhelming, will
place limitations on the reliability of the LT separation technique in
extracting high-$Q^2$ form factors. On the other hand, the size of the
2$\gamma$ contributions to elastic scattering could be determined from
measurement of the ratio of $e^-p$ to $e^+p$ elastic cross sections,
which are uniquely sensitive to 2$\gamma$ exchange effects.

We have also generalized our analysis to the case where the initial
electron and recoil proton are polarized, as in the polarization
transfer experiments. While the 2$\gamma$ corrections can be as large as
$\sim 4$--5\% at small $\varepsilon$ for $Q^2 \sim 6$~GeV$^2$, because
the polarization transfer measurements are performed typically at large
$\varepsilon$ we find the impact on the extracted $G_E^p/G_M^p$ ratio to
be quite small, amounting to $\alt 3\%$ suppression at the highest $Q^2$
value.

Extending the formalism to the case of the neutron, we have calculated
the 2$\gamma$ exchange corrections to the neutron $G_E^n/G_M^n$ ratio.
While numerically smaller than for the proton, the corrections are
nonetheless important since the magnitude of $G_E^n$ itself is small
compared with $G_E^p$. Furthermore, because of the opposite sign of the
neutron magnetic moment relative to the proton, the 2$\gamma$
corrections to the LT-separated cross section give rise to a sizable
enhancement of $G_E^n/G_M^n$ at large $Q^2$. The analogous effects for
the polarization transfer ratio are small, on the other hand, giving
rise to a few percent suppression for $Q^2 \alt 6$~GeV$^2$.

Finally, we have also obtained an estimate of the 2$\gamma$ exchange
contribution to the elastic form factors of $^3$He from elastic
intermediate states. The results reveal an interesting interplay between
an enhancement from the larger charge of the $^3$He nucleus and a
suppression due to the larger mass. Together with softer form factor
(larger radius) compared with that of the nucleon, the net effect is
$\alt 1\%$ over the $Q^2$ range accessible to current and upcoming
experiments.

Contributions from excited states, such as the $\Delta$ and heavier
baryons, may modify the quantitative analysis presented here. Naively,
one could expect their effect to be suppressed because of the larger
masses involved, at least for the real parts of the form factors.
An investigation of the inelastic excitation effects is presented in 
Ref.~\cite{Kon05}.

\acknowledgments

We would like to thank J.~Arrington for helpful discussions and 
communications. This work was supported in part by NSERC
(Canada), DOE grant DE-FG02-93ER-40762, and DOE contract
DE-AC05-84ER-40150 under which the Southeastern Universities Research
Association (SURA) operates the Thomas Jefferson National Accelerator
Facility (Jefferson Lab).


\newpage

\begin{table}
\begin{center}
\caption{Parameters for the proton and neutron form factor fits in
	Eq.~(\protect\ref{eq:3pole}) used in this work, with $n_i$ and
	$d_i$ in units of GeV$^2$.}
\vspace*{0.5cm}
\begin{tabular}{|c|rr|rr|}				\hline
	& $F_1^p$\ \ \ \ \ & $F_2^p$\ \ \ \ \
	& $F_1^n$\ \ \ \ \ & $F_2^n$\ \ \ \ \		\\ \hline
\ $N$\	& 3\ \ \ \ \ & 3\ \ \ \ \ & 3\ \ \ \ \ & 2\ \ \ \ \  \\ \hline
\ $n_1$\ &\ 0.38676\ & 1.01650\  & 24.8109\   &\ 5.37640\  \\
\ $n_2$\ &\ 0.53222\ & --19.0246\ & --99.8420\ &            \\
\ $d_1$\ &\ 3.29899\ & 0.40886\  & 1.98524\   &\ 0.76533\  \\
\ $d_2$\ &\ 0.45614\ & 2.94311\  & 1.72105\   &\ 0.59289\  \\
\ $d_3$\ &\ 3.32682\ & 3.12550\  & 1.64902\   & ---\ \ \ \ \\
\hline
\end{tabular}
\label{tab:param}
\end{center}
\end{table}

\end{document}